\documentclass[useAMS,usegraphicx]{mn2e}
\usepackage{times}
\usepackage{rotate}
\usepackage{amssymb}
\title[X-ray selected starbursts in the GOODS-North]{X-ray selected starbursts in the GOODS-North} 
\author[Georgakakis et al. ] {A. Georgakakis$^{1}$\thanks{Marie-Curie
    fellow; email: age@imperial.ac.uk}, M. Rowan-Robinson$^{1}$, 
  T. S. R. Babbedge$^{1}$,  
  I. Georgantopoulos$^2$  
  \\ \\
  $^1$Astrophysics Group, Blackett Laboratory, Imperial College, Prince
  Consort Rd , London SW7 2AZ, UK\\
  $^2$National Observatory of Athens, V. Paulou \& I. Metaxa, 11532,
  Greece\\   
 }
\begin{document}
\maketitle  

\begin{abstract}  We investigate claims  that recent  ultra-deep X-ray
surveys  are   detecting  starbursts  at   cosmologically  interesting
redshifts  ($z=0-1$).   We combine  X-ray  data  from  the 2\,Ms  {\it
Chandra} Deep  Field North and  multi-wavelength observations obtained
as part of the GOODS-North  to build the Spectral Energy Distributions
(UV, optical, infrared) of X-ray sources in this field.  These are fit
with  model templates  providing  an estimate  of  the total  infrared
luminosity ($\rm  3-1000\mu m$) of  each source.  We then  exploit the
tight  correlation   between  infrared  and   X-ray  luminosities  for
star-forming galaxies,  established in  the local Universe,  to select
sources that are dominated  by star-formation rather than supermassive
black  hole accretion.  This approach  is efficient  in discriminating
normal galaxies  from AGN over  a wide range of  star-formation rates,
from quiescent systems to starbursts. The above methodology results in
a  sample  of 45  X-ray  selected  star-forming  systems at  a  median
redshift  $z\approx0.5$,  the  majority  of which  (60\%)  are  either
Luminous or  Ultra-Luminous Infrared  Galaxies.  This sample  is least
affected  by  incompleteness and  residual  AGN  contamination and  is
therefore well suited for cosmological studies.  We quantify the X-ray
evolution of  these sources  by constructing their  differential X-ray
counts, dN/dS,  and comparing  them with evolving  luminosity function
models.  The  results are consistent with luminosity  evolution of the
form  $(1+z)^{p}$  with $p  \approx  2.4$.   This  is similar  to  the
evolution rate of star-forming galaxies selected at other wavelengths,
suggesting that  the deep X-ray  surveys, like the {\it  Chandra} Deep
Field North,  are indeed finding the starburst  galaxy population that
drives the  rapid evolution of the global  star-formation rate density
in the  range $z  \approx 0-1$. Our  analysis also reveals  a separate
population  of infrared-faint  X-ray sources  at  moderate-$z$.  These
include  old galaxies  but also  systems that  are X-ray  luminous for
their stellar mass compared to local ellipticals.  We argue that these
may be post-starbursts  that will, over time, become  fainter at X-ray
wavelengths and  will eventually  evolve into early-type  systems (i.e
E/S0).
\end{abstract}

\section{Introduction}\label{sec_intro} 

In the  last two decades independent  observational programs selecting
galaxies over almost the  entire range of the electromagnetic spectrum
(UV, optical, infrared, sub-mm,  radio) have consistently identified a
fast evolving population of actively star-forming galaxies at moderate
and high-$z$ (e.g.  Rowan-Robinson 1993; Lilly et al.  1996; Wilson et
al. 2002;  Takeutchi et al. 2003;  Babbedge et al.   2006; see Hopkins
2004 for  a more  complete reference list).   These systems  drive the
rapid  evolution of the  global star-formation  rate density  from the
local Universe to $z \approx 1 - 2$ (e.g.  Hopkins 2004).

Recently,  surveys with  the {\it  Chandra} and  the  {\it XMM-Newton}
missions  have claimed  the detection,  for  the first  time at  X-ray
wavelengths,  of  this star-forming  galaxy  population. For  example,
stacking  the X-ray  photons at  the positions  of  optically selected
galaxies, not  individually detected  at X-rays, provide  estimates of
the mean X-ray properties of star-forming systems out to $z \approx 3$
(Brandt et al.  2001; Hornschemeier et al. 2002;  Nandra et al.  2002;
Georgakakis et al.  2003; Laird  et al.  2005, 2006). Hornschemeier et
al.   (2000) reported  one of  the  first direct  detections of  X-ray
emission  from  a  normal  galaxy  outside the  local  Universe:  an
off-nuclear  X-ray source  associated  with a  low-$z$  spiral in  the
HDF-N, most likely an X-ray  binary or a supernova remnant.  Alexander
et al.  (2002)  find a large overlap between $z  \la 1$ narrow optical
emission-line  X-ray sources  in the  1\,Ms Chandra  Deep  Field North
(CDF-North)  and ISO  $\rm 15\,\mu  m$ detections  in the  Hubble Deep
Field North  (HDF-N).  The X-ray and mid-infrared  properties of these
systems are found  to be consistent with starburst  activity. Bauer et
al.   (2002)  using  data from  the  same  field  also report  a  high
identification  rate  between  $\mu$Jy  radio  sources,  dominated  by
star-forming   galaxies,  and   narrow  optical   emission-line  X-ray
detections.   The radio and  X-ray luminosities  of these  systems are
also  consistent with star-formation  (but see  Barger et  al.  2007),
further suggesting  the appearance  of $z \la  1$ starbursts  at X-ray
wavelengths.  Quiescent early  and late-type galaxies at $z\approx0.3$
have also been identified in  the 2\,Ms CDF-North by selecting sources
with low X-ray--to--optical flux ratios ($\log f_X / f_{opt} \la -2$),
about 2\,dex below typical  AGN (Hornschemeir et al.  2003).  Parallel
to  these ultra-deep  pencil-beam studies,  shallow  wide-area samples
with  both  the  {\it  Chandra}  and the  {\it  XMM-Newton}  are  also
routinely   finding  normal  galaxy   candidates  at   lower  redshift
$z\approx0.1$ (e.g. Georgakakis et al.  2006a; Kim et al.  2006).  The
number  counts of this  new (for  X-ray surveys)  population increases
steeply with decreasing X-ray  flux, suggesting that it will outnumber
AGN  below  the  flux  limit  of the  current  deepest  surveys  (e.g.
Horschemeier  et al.  2003;  Bauer et  al.  2004;  Kim et  al.  2006).
These sources will therefore be the dominant component of future X-ray
samples.

In  addition to  the observational  work  above, studies  are also  in
progress  to model the  physical processes  responsible for  the X-ray
emission of  star-forming galaxies  (e.g.  accreting binaries)  and to
understand  how  these systems  evolve  with  time (e.g.   Belczynski,
Kalogera \& Bulik  2002; Sipior 2003; Belczynski et  al. 2007).  Ghosh
\& White (2001)  adopt a semi-empirical approach to  link the lifetime
of accreting  binaries with the  star-formation rate and to  place the
X-ray evolution  of galaxies in a cosmological  context.  Depending on
the  star-formation   history  of  the  Universe   and  the  evolution
timescales of the high and  low mass X-ray binaries (LMXRBs and HMXRBs
respectively),  which   dominate  the  X-ray   emission  of  late-type
galaxies,  these authors  predict different  evolution rates  at X-ray
wavelengths  compared to other  wavebands.  Inverting  this argument,
constraining the  X-ray evolution of galaxies  can potentially provide
information  on  physical   properties,  such  as  the  characteristic
timescales of  LMXRBs and HMXRBs.  For example, if HMXRB  dominate the
X-ray emission, we expect the total X-ray luminosity to follow closely
the star-formation history of the  Universe as observed at optical and
IR wavelengths.  In contrast, the LMXRB evolve much more slowly and in
this case the integrated X-ray luminosity of galaxies should present a
significant  time  lag behind  the  optical/IR star-formation  density
profile.   The number  density and  evolution of  LMXRB  also presents
interest  for  the  LISA  gravitational  wave mission.  This  will  be
sensitive to the gravitational wave radiation of binaries with orbital
periods  less than  4\,hr.   These sources  will  consist primarily  of
LMXRB,  where the  companion is  a  low mass  ($\rm 0.4M_\odot$)  main
sequence  star (Verbunt \&  Nelemans 2001).  The determination  of the
LMXRB evolution  with cosmic time  holds important information  on the
number of  expected LISA  detections as  well as on  the level  of its
background radiation

The evidence above has motivated attempts to constrain observationally
the X-ray evolution of these systems. Norman et al.  (2004) compiled a
sample  of normal  galaxy candidates  in the  CDF-North and  South and
determined  their luminosity  function at  different redshifts  in the
range $z\approx0-1$.  Their analysis indicates luminosity evolution of
the  form  $\approx(1+z)^{2.7}$, consistent  with  that inferred  from
other wavebands.   Similar results  are obtained by  the complementary
approach  of constructing  the  X-ray number  counts  of galaxies  and
comparing  them  with   evolutionary  model  predictions  (Ranalli  et
al. 2006; Georgakakis et al.  2006b).  The evidence above implies that
the  galaxy populations  in X-ray  surveys are  dominated by  the fast
evolving starbursts,  also found at other wavelengths,  and that their
X-ray emission is predominantly  from short-lived HMXRB that trace the
current star-formation rate of the host galaxy.

However,  AGN   remain  the   dominant  component  of   current  X-ray
surveys. As  a result  the nature of  the X-ray  selected star-forming
galaxy   candidates  identified  in   recent  X-ray   samples  remains
controversial.  Although  star-formation is an  attractive option, the
role of low-level or obscured AGN activity, that may dominate at X-ray
wavelengths,   is  unclear.    As  in   any  waveband,   residual  AGN
contamination  in  X-ray  selected  star-forming  galaxy  samples  may
significantly  distort  any  conclusions  on the  evolution  of  these
systems,  casting doubt  on the  results  of the  studies above.   For
example, Bauer  et al.  (2004) use  a range of  criteria (e.g. optical
and X-ray spectra, X-ray--to--optical flux ratio, X-ray luminosity) to
group   X-ray    sources   in    the   CDF-North   and    South   into
obscured/unobscured AGN  and normal  galaxy candidates. The  number of
sources in  the latter  sample varies by  about 50\% depending  on how
conservative  the   criteria  are   in  excluding  any   possible  AGN
contribution. Therefore, any attempt  to constrain the X-ray evolution
of  star-forming galaxies  strongly  depends on  the sample  selection
(e.g.  Georgakakis  et al.  2006b). In  order to place  this new X-ray
population in a cosmological context  it is important that we quantify
the significance  of star-formation and  AGN activity to  the observed
X-ray emission of these sources.

In this  paper we combine deep  X-ray, optical and mid-IR  data in the
Great  Observatories  Origins  Deep  Survey (GOODS)  North  region  to
construct the  Spectral Energy Distributions (SEDs)  of X-ray selected
starburst  candidates  and  to  constrain  their  dominant  energising
source, star-formation  or AGN activity. We exploit  in particular the
tight  correlation   between  X-ray  and   infrared  luminosities  for
star-forming galaxies (e.g. Ranalli et  al.  2003) to compile an X-ray
selected   sample   of   such   systems,   least   affected   by   AGN
contamination.  The advantage  of  this  approach is  that  it is  not
affected by biases introduced  when selecting star-forming galaxies by
applying  cuts  in  X-ray--to--optical  flux ratio  and/or  the  X-ray
luminosity. Our main goal is  to explore the cosmological evolution of
starbursts at X-ray wavelengths.   Throughout this paper we adopt $\rm
Ho = 72 \, km \, s^{-1}  \, Mpc^{-1}$, $\rm \Omega_{M} = 0.3$ and $\rm
\Omega_{\Lambda} = 0.7$.

\section{Data}\label{sec_data}

Deep  multi-waveband optical  imaging  ($UBVRIz$) in  the GOODS  North
region has been obtained by  Capak et al.  (2004). The data reduction,
source detection  and catalogue generation are described  in detail by
these authors.   These observations cover about $\rm  0.2\, deg^2$ and
extend beyond the GOODS field  of view ($\approx \rm 0.05\,deg^2$). In
this  study we  use the  $R$-band selected  sample  comprising 47\,451
sources to the limit $R_{AB}=26.6$\,mag ($5\sigma$).

X-ray  observations  are available  as  part  of  the 2\,Ms  CDF-North
survey. This  covers a total of  about $\rm 17 \times  17 \, arcmin^2$
and provides the deepest X-ray  sample currently available [$f_X ( \rm
0.5-2  \,  keV)  \approx 2.5  \times  10^{-17}  \,  erg \,  s^{-1}  \,
cm^{-2}$].   In this  paper we  use the  X-ray point  source catalogue
constructed by Alexander et al. (2003).

The mid-infrared  is one of the  key wavelength regimes  of the GOODS.
The observations were carried out by the Spitzer Space Telescope using
the IRAC (3.6, 4.5, 5.8 and 8.0$\, \rm \mu m$) and MIPS (24$\, \rm \mu
m$) detectors. These cover a total area of about $10 \times 16.5\rm \,
arcmin^2$ on the  sky, smaller than the CDF-North  but centered on the
most sensitive  region of the X-ray observations.   The data reduction
is based on the Spitzer Science Center pipeline and is further refined
by  the  GOODS  team  using  custom routines  (Dickinson  et  al.   in
preparation).   The final data  products comprise  photometrically and
astrometrically calibrated mosaiced images  in all five wavebands.  In
this paper  we use the  2nd data release  (DR2) of the  IRAC superdeep
images (version 0.30) and the  interim data release (DR1+) of the MIPS
24$\,  \rm \mu m$  mosaic (version  0.36) provided  by the  GOODS team
(Dickinson  et  al.  2003;  Dickinson  et  al.   in preparation).  The
pixelscale  is  0.6  and  1.2\,arcsec  for  the  IRAC  and  MIPS  data
respectively.  Sources  are detected on  the individual IRAC  and MIPS
wavebands  by  running SExtractor  (Bertin  \&  Arnouts  1996) on  the
exposure-map  weighted  images   convolved  with  a  $5\times5$  pixel
Gaussian  filter with  a full-width  half  maximum (FWHM)  of 1.2  and
3\,arcsec for the IRAC  and MIPS observations respectively. The source
fluxes are estimated using apertures with diameters of 3.6\,arcsec for
the IRAC bands  and 7.2\,arcsec for the MIPS  image.  Total fluxes are
estimated by multiplying the  measured fluxes with aperture correction
factors. These are determined  by integrating the Spitzer point spread
function (PSF) in different wavebands  and are found to be 1.40, 1.47,
1.73, 1.86 and 2.49 for the IRAC  3.6, 4.5, 5.8 and 8.0 and MIPS 24$\,
\rm \mu m$ bands respectively.  In the case of non-detection we assign
an upper limit to the flux which corresponds to a value close the peak
of the source counts in individual wavebands.

\section{The Sample}\label{sec_sample}

We choose  to work in the  soft spectral band  (0.5-2\,keV), using the
CDF-N  catalogue of  Alexander et  al. (2003),  as the  sensitivity of
Chandra  is highest in  this energy  range and  also because  late and
early-type  galaxies  typically have  soft  X-ray spectral  properties
($\Gamma \approx 2.0$, $\rm kT = 0.5-2$\,keV; e.g Fabbiano 1995).  For
the initial  classification of the X-ray sources  into different types
we use  the scheme  presented by  Bauer et al.   (2004) as  a starting
point.  At this stage we do  not use the IR information.  Bauer et al.
(2004) grouped  X-ray sources  into galaxy candidates  and AGN  on the
basis of their  (i) X-ray spectra, (ii) radio  properties, (iii) X-ray
luminosity, (iv)  optical spectroscopy when  available.  In particular
galaxies have off-nuclear X-ray emission, soft X-ray spectra with $N_H
< \rm  10^{22} \,cm^{-2}$ or  hardness ratio $<0.8$,  radio properties
that do  not suggest AGN  activity, $L_X(  \rm 0.5 -  8.0 \, keV)  < 3
\times  10^{42} \,  erg \,  s^{-1}$, somewhat  brighter than  the most
X-ray   luminous   starbursts    in   the   local   Universe   (Zezas,
Georgantopoulos \& Ward 1998; Moran,  Lehnert \& Helfand 1999) and low
X-ray  to optical  flux ratio,  $\log f_X  / f_R<-1$.   We  define our
``secure'' galaxy sample  to include all the sources  that fulfill the
criteria above.  This conservatively  selected sample is least affected
by  AGN contamination  at  the expense  of  potentially missing  X-ray
luminous starbursts and/or massive ellipticals (i.e.  $L_X \ga 10^{42}
\rm \, erg \, s^{-1}$; e.g. Tzanavaris, Georgantopoulos \& Georgakakis
2006).   We address this  issue by  considering as  potential galaxies
X-ray sources  that although  classified AGN by  Bauer et  al.  (2004;
e.g.   high $L_X$,  $\log  f_X  / f_R>-1$)  have  soft X-ray  spectral
properties (i.e.  not obscured AGN) and optical emission that does not
reveal AGN activity (i.e. broad emission lines). Most of these systems
are Seyferts but luminous starbursts with  $L_X( \rm 0.5 - 8.0 \, keV)
\ga 3  \times 10^{42}  \, erg  \, s^{-1}$ may  also be  present. These
sources define  our ``optimistic'' galaxy  sample.  Late-type galaxies
likely exist  in both the  ``optimistic'' and the  ``secure'' samples.
For clarity  we keep these  two sub-samples independent  and therefore
there is no overlap between the  sources in each of them. In section 5
we  include the the  IR information  in the  analysis to  improve this
classification by minimising the AGN contamination of the star-forming
galaxy sample.

The  X-ray detections are  identified with  optical sources  using the
photometric catalogues of Capak et  al. (2004) with a search radius of
3\,arcsec. The  optical counterparts are identical  to those presented
by Barger  et al.  (2003).  For sources  with optical identifications,
we use the optical centroid,  which has better positional accuracy, to
cross-correlate  with  the IRAC  and  MIPS  source positions.   Visual
inspection confirms that the  mid-IR counterparts are robust. A number
of X-ray sources lie outside the IRAC or MIPS field of view. These are
excluded from the analysis that  follows. Our final sample comprises a
total of 230 0.5-2\,keV  selected sources with optical identifications
that  overlap with  both  the IRAC  and  the MIPS  survey regions  and
therefore have detections or upper limits in all bands, $\rm 3.6-24\mu
m$.  This includes 56 and 59 independent sources in the ``optimistic''
and ``secure'' samples respectively.  There are additionally 95 and 20
sources  classified  obsucred (type-II)  and  unobscured (type-I)  AGN
respectively by Bauer et al. (2004).  These systems are not considered
as potential galaxy candidates but will serve as comparison sample.

\section{Spectral Energy Distributions}\label{sec_sed}

The observed  optical to mid-IR Spectral Energy  Distribution (SED) of
the  X-ray sources  in our  sample are  modeled following  the methods
fully described in Rowan-Robinson et al. (2005). In brief the $U$-band
to $\rm  4.5\,\mu m$  photometric data  are fit using  a library  of 8
templates described  by Babbedge  et al. (2004),  6 galaxies  (E, Sab,
Sbc, Scd, Sdm and sb) and 2  AGN. At longer wavelengths ($\rm 5.8 - 24
\, \mu m$) any dust  may significantly contribute or even dominate the
observed  emission. Before  fitting  models to  these wavelengths  the
stellar  contribution  is  subtracted  from the  photometric  data  by
extrapolating  the best-fit  galaxy template  from the  previous step.
The residuals  are then fit with  a mixture of  four templates: cirrus
(Efstathiou  \& Rowan-Robinson  2003), AGN  dust  tori (Rowan-Robinson
1995;   Efstathiou  \&  Rowan-Robinson   1995),  M\,82   and  Arp\,220
starbursts (Efstathiou  et al.  2000). The modeling  above provides an
estimate of  the total infrared  luminosity, $L_{TOT}$, of  our sample
sources in  the wavelength  range $\rm 3-1000\mu  m$. As  discussed by
Rowan-Robinson et al. (2005) this  is expected to be accurate within a
factor of two.  For sources  without spectroscopic redshift we use the
photometric redshift  estimated from the optical  template SED fitting
to determine  $L_{TOT}$. The accuracy of the  photometric redshifts is
estimated $\delta z/(1+z_{spec})\approx 0.04$ ($1\sigma$ rms). We note
however,  that  most  of  the  galaxy  candidates  have  spectroscopic
redshifts available. A total of 12 and 7 sources in the ``secure'' and
``optimistic'' samples respectively show  no infrared excess above the
stellar expectation and therefore $L_{TOT}$ cannot be estimated. These
infrared-faint  sources are  most likely  early-type galaxies  and are
discussed  separately.   There  are  47  and  49  star-forming  galaxy
candidates in the  ``secure'' and ``optimistic'' samples respectively,
which show mid-infrared excess above the stellar prediction.

In addition  to the infrared luminosities above  we also independently
estimate stellar masses ($M_\star$) for the GOODS-North X-ray sources.
Our method is similar to those used by other groups (e.g. Bundy et al.
2005) and  is based on $\chi^2$  minimisation to find, from  a grid of
template  SEDs,  the  best-fit  solution to  the  observed  broad-band
photometry  of each  galaxy ($UBVRIz$  and IRAC  3.6$\mu$m).   The SED
library  consists  of a  large  number  (about  20\,000) of  synthetic
spectra  extracted  using  the   Bruzual  \&  Charlot  (2003)  stellar
population synthesis  code adopting  the Chabrier (2003)  Initial Mass
Function  (IMF).  The  SEDs span  a range  of  exponentially declining
star-formation  histories (e-folding  time  $\tau=0.1-15$\,Gyrs), ages
since     the     initial     burst    ($<15$\,Gyr),     metallicities
($0.002-2.5\,Z_{\odot}$)   and   diffuse   interstellar  medium   dust
obscuration ($A_V=0-0.5$) using the  SMC extinction law (Gordon et al.
2003).  When  comparing model  with observation we  only use  the IRAC
3.6$\mu$m photometry  from the  available Spitzer bands.   The stellar
emission decreases significantly at  longer wavelengths.  Also, in the
fitting  process  the  redshift  of  source is  fixed  to  either  the
spectroscopically  determined   value  (when  available)   or  to  the
photometric  redshift estimated  above. Stellar  masses  are estimated
using the mass-to-light ratio ($M/L$) at 3.6$\mu$m of the best-fit SED
and  the observed  luminosity of  the  galaxy at  this waveband.   The
advantage of using this wavelength is that first, it is least affected
by extinction and  second, it lies close to  the stellar emission peak
of the galaxy SED proving a reliable census of the stellar mass of the
system.   We note  that although  the allowed  template  SED parameter
space is large, a number  of uncertainties and systematic effects such
as simplistic  assumptions about the adopted SEDs  (e.g.  single burst
stellar populations) and aliases  between different templates may bias
our  results.   Bundy  et  al.   (2005) quantify  the  uncertainty  in
$M_\star$ because of these effects.  Based on this study we expect the
derived stellar  masses to  be accurate within  a factor of  two.  The
observed and  rest-frame properties  of sources in  the ``optimistic''
and  ``secure''  samples are  listed  in  Table  1 for  infrared-faint
sources and Table 2 for the star-forming galaxy candidates.

\section{Results}\label{sec_results}

\begin{figure}
 \rotatebox{0}{\includegraphics[height=1\columnwidth]{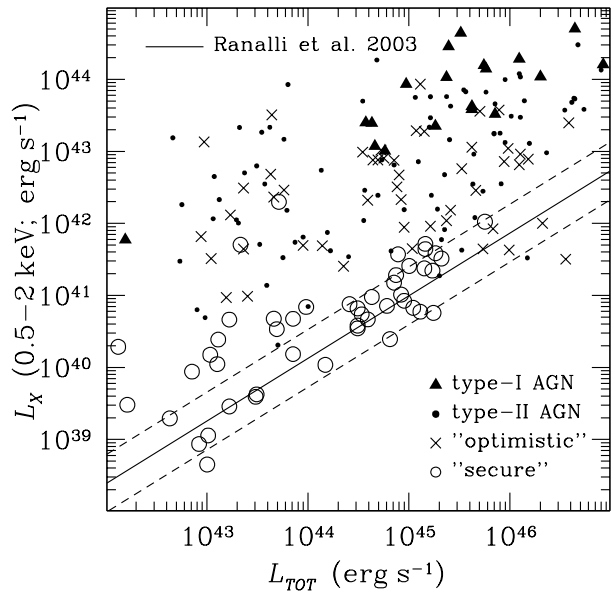}}
\caption{
 0.5-2\,keV X-ray luminosity as a function of total infrared
 luminosity, $L_{TOT}$. The open circles and the crosses correspond to
 the ``secure'' and ``optimistic'' galaxy samples respectively described in
 section \ref{sec_sample}. The filled circles and
 triangles are X-ray sources classified type-II AGN and broad-line
 type-I AGN respectively by Bauer et al. (2004). The continuous line
 is the X-ray/infrared luminosity relation for local star-forming
 galaxies from Ranalli et al. (2003). The dashed lines correspond to
 the $2\sigma$ rms envelope around this relation and define the
 selection box for the ``normal galaxy'' sample. The $1\sigma$ rms
 is estimated by taking into account both the scatter in the Ranalli
 et al. data and the factor of 
 2 uncertainty in the $L_{TOT}$ determination.
 } \label{fig_lx_lir}
\end{figure}

\begin{figure}
 \rotatebox{0}{\includegraphics[height=1\columnwidth]{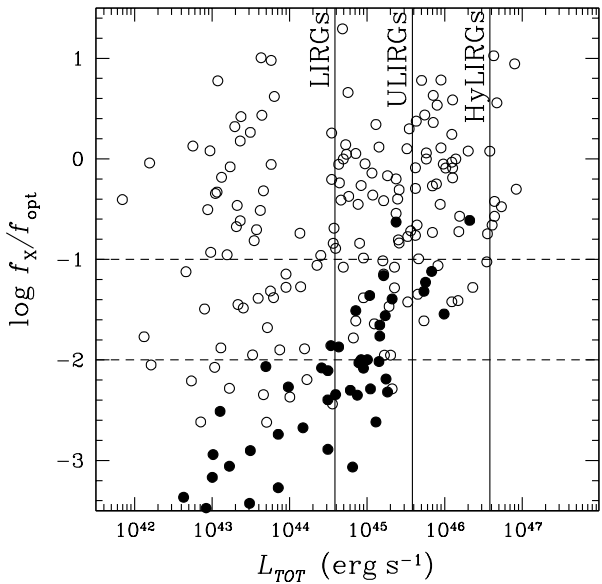}}
\caption{ 
Rest-frame  X-ray--to--optical  flux   ratio  against  total  infrared
luminosity, $L_{TOT}$.  The open circles correspond to AGN dominated
X-ray  sources   that  deviate  by   more  than  $2\sigma$   from  the
X-ray/infrared  luminosity relation  for local  star-forming galaxies.
Filled circles  are X-ray selected  ``normal galaxies'', i.e. X-ray
sources that scatter $<2\sigma$ from this relation and are most likely
dominated  by star-formation.  The  horizontal dashed  lines mark  the
limits  $\log f_X/f_{opt}=-1$ and  $-2$. The  vertical lines  show the
$L_{TOT}$ limits for luminous (LIRGs;  $ > \rm 10^{11} \, L_{\odot}$),
ultra-luminous   (ULIRGs;  $   >  \rm   10^{12}  \,   L_{\odot}$)  and
hyper-luminous  (HyIRGs;  $  >  \rm 10^{13}  \,  L_{\odot}$)  infrared
galaxies.  
}\label{fig_fxfopt} 
\end{figure}

In Figure \ref{fig_lx_lir} we  plot X-ray luminosity in the 0.5-2\,keV
spectral  band, after  correcting for  intrinsic  obscuration, against
$L_{TOT}$  for  the  GOODS-N   X-ray  sources,  both  AGN  and  galaxy
candidates. For comparison we  also show the X-ray/infrared luminosity
relation  for the sample  of local  star-forming galaxies  compiled by
Ranalli  et  al.  (2003).   These  authors  showed  that for  galaxies
dominated by  star-formation the  X-ray and infrared  luminosities are
tightly correlated, while AGNs deviate from this relation in the sense
that  they are  X-ray luminous  for their  infrared  luminosity.  This
relation can therefore be used to discriminate between AGN and systems
dominated  by star-formation.  In  order to  compare our  results with
those of Ranalli et al.   (2003) we convert their infrared luminosity,
$L_{IR}$, estimated using the IRAS  60 and $\rm 100\mu m$ bands (Helou
et  al. 1988),  to  total infrared  luminosity  used here,  $L_{TOT}$,
adopting  the  conversion  $L_{IR}=0.7  \times L_{TOT}$  (Takeuchi  et
al. 2005). We also estimate the expected $1\sigma$ rms envelope of the
X-ray/infrared  luminosity  relation   for  star-forming  galaxies  by
quadratically  adding   the  $1\sigma$  scatter  of   the  Ranalli  et
al. relation ($\sigma=0.27$) and the factor of 2 ($\delta \log \approx
0.3$) uncertainty in $L_{TOT}$.

In Figure  \ref{fig_lx_lir} the  majority of X-ray  sources classified
AGN by Bauer et al. (2004) deviate from the $L_X-L_{TOT}$ relation for
star-forming  galaxies. Also  most of  the sources  in  the ``secure''
sample scatter  around the Ranalli et al.   (2003) relation suggesting
that  their  X-ray  emission   is  dominated  by  star-formation.   As
expected, the AGN contamination of the ``optimistic'' sample is large,
in agreement with the classification of Bauer et al. (2004). There are
nevertheless  few sources  in a  that  subsample that  have X-ray  and
infrared  luminosities consistent  with star-formation,  notably those
with  high  $L_{TOT}$.   In  the  discussion that  follows  we  define
``normal  galaxies''  those  X-ray   sources  in  the  ``secure''  and
``optimistic'' samples  which deviate by less than  $2\sigma$ from the
Ranalli et  al.  (2003)  relation. Galaxies selected  in this  way are
assumed to  be dominated by star-formation.  Relaxing  the above limit
for  ``normal  galaxy'' selection  will  not  significantly alter  our
conclusions but  is likely  to include an  increasing fraction  of AGN
dominated  systems in  the  ``normal galaxy''  sample.   Based on  the
above,  the fraction  of  ``normal galaxies''  in  the ``secure''  and
``optimistic''  samples  (excluding  infrared  faint  sources)  is  80
(38/47) and 14 (7/49) per cent respectively.

Also  in Figure \ref{fig_lx_lir}  we find  only one  ``normal galaxy''
with X-ray luminosity marginally in excess of $L_X(\rm 0.5 - 2 \, keV)
= 10^{42} \, erg \, s^{-1}$.  The lack of X-ray luminous starbursts in
the sample  is independent of  the Bauer et al.  (2004) classification
and  is  based  only  on  the  distance  of  X-ray  sources  from  the
$L_X-L_{TOT}$  relation  of  star-forming  galaxies.   In  the  nearby
Universe  the X-ray  brightest  starbursts known  also  have $L_X  \la
10^{42} \rm  \, erg \,  s^{-1}$ (e.g.  Zezas, Georgantopoulos  \& Ward
1998;  Moran, Lehnert  \&  Helfand 1999)  suggesting  that this  X-ray
luminosity may correspond  to the upper limit that  can be produced by
star-formation.   The evidence  above justifies  the use  of  an upper
limit in $L_X$ when  searching for star-formation dominated sources in
X-ray samples (e.g.  Bauer et al. 2004; Tzanavaris  et al. 2006). This
issue will be discussed in more detail in section 6.

In addition  to the $L_X$  cutoff, the X-ray--to--optical  flux ratio,
$\log  f_X  /f_{opt}$,  is  also  used to  select  against  AGN
(e.g. Hornschemeier  et al.   2003; Georgantopoulos  et al.
2005). Our ``normal galaxy'' sample is independent of $\log  f_X
/f_{opt}$ and therefore, can be used to assess the level of bias
introduced when selecting star-forming systems using this quantity.
Figure \ref{fig_fxfopt}  plots $\log  f_X /f_{opt}$  against
$L_{TOT}$.  The X-ray--to--optical  flux   ratio  is  estimated
using  the  $R$-band magnitude  and  the 0.5-2\,keV  flux  (not
corrected for  absorption) according to the relation (Hornschemeier et
al. 2003) 

\begin{equation}\label{eq1}     \log\frac{f_X}{f_{opt}}     =     \log
f(0.5-2\,{\rm keV}) + 0.4\,R + 5.5.
\end{equation}

\noindent  Rather  than using  rest-frame  fluxes  for  the $\log  f_X
/f_{opt}$ we prefer to use observed quantities because this is what is
available  to  observers  when  selecting  their  sample. In  Figure
\ref{fig_fxfopt}  the   X-ray--to--optical  flux  ratio   of  ``normal
galaxies'' increases,  on average, with  $L_{TOT}$.  Quiescent systems
with low-level star-formation rate  have $\log f_X /f_{opt}<-2$, while
more active  galaxies have $-2<\log f_X  /f_{opt}<-1$ (e.g.  Alexander
et al.   2003).  AGN dominated  systems although typically  have $\log
f_X /f_{opt}>-1$, also overlap with  the region of the parameter space
occupied  by starbursts  resulting in  significant  contamination when
selecting  galaxies only  on the  basis of  $\log f_X  /f_{opt}$.  For
example, in the case of  $\log f_X /f_{opt}<-1$, about 53\% (48/91) of
the GOODS-North X-ray sources  are AGN dominated.  This underlines the
need  for multiwavelength  data  to identify  star-forming systems  in
X-ray  selected samples.  Selecting  at lower  X-ray--to--optical flux
ratios,   i.e.   $\log   f_X  /f_{opt}<-2$,   though   minimising  AGN
contamination (28\%; 11/39), results  in incompleteness since the most
X-ray luminous starbursts are not included. Georgakakis et al. (2006b)
show that it is precisely those sources with $-2 < \log f_X /f_{opt} <
-1$, that drive  the evolution of the X-ray  population.  We also note
that this incompleteness  is likely to be more  severe at moderate and
high-$z$ samples  since at lower redshifts  ($z\approx0.1$) such X-ray
luminous starbursts are rarer.  For example using the late-type galaxy
X-ray  luminosity  function  of   Georgakakis  et  al.   (2006a)  with
luminosity evolution of  the form $(1+z)^{3}$ we find  that the volume
density of sources with $L_X  > 10^{40}\rm \, erg \, s^{-1}$ decreases
by a  factor of about 2.5  from $z\approx0.5$ (the  median redshift of
the ``normal galaxy'' sample presented here) to $z\approx0.1$.

Finally, in  Figure \ref{fig_fxfopt}  we find that  60 (27/45)  and 11
(5/45) per  cent of the  X-ray selected ``normal galaxies''  belong to
the luminous (LIRGs; $L_{TOT} > 5 \times 10^{44}\rm \, erg \, s^{-1}$)
and ultra-luminous (ULIRGs;  $L_{TOT} > 5 \times 10^{45}\rm  \, erg \,
s^{-1}$)  infrared  galaxy   classes  respectively.   This  is  direct
evidence that  current ultra-deep  X-ray surveys detect  the starburst
galaxy population that  is responsible for the rapid  evolution of the
global star-formation rate from the local Universe to $z \approx 1-2$.

\begin{figure}
 \rotatebox{0}{\includegraphics[height=1\columnwidth]{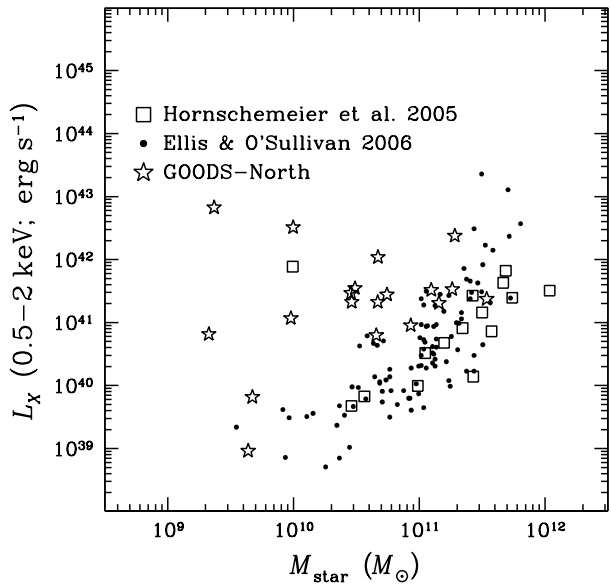}}
\caption{
 Stellar mass against X-ray luminosity for sources with SEDs that do
 not show infrared excess above the stellar prediction. Also plotted
 for comparison are the local ellipticals from the sample of Ellis \&
 O'Sullivan (2006; for clarity we do not plot upper limits) and the
 absorption-line sources from the  SDSS sample of Hornschemeier et
 al. (2005). 
 }\label{fig_lxmass}
\end{figure}

Next  we   consider  those  X-ray   sources  in  the   ``secure''  and
``optimistic'' samples with SEDs that do not show infrared excess over
the stellar prediction. For these sources we cannot estimate $L_{TOT}$
and  therefore  they are  not  shown in  any  of  the plots  presented
above. Many of  them (but not all) have  early-type optical morphology
in  the ACS  survey  of  the GOODS-North  or  absorption line  optical
spectra (see below). These systems are either early type galaxies with
little star-formation or weak  AGN.  Figure \ref{fig_lxmass} plots the
stellar mass of these sources as a function of their X-ray luminosity.
For  comparison we  also show  nearby ellipticals  from the  sample of
Ellis   \&   O'Sullivan   (2006)   as  well   as   the   $z\approx0.1$
absorption-line  SDSS  galaxies  presented  by  Hornschemeier  et  al.
(2005). For  the former sample  stellar masses are  approximated using
the $K$-band magnitudes  listed by Ellis \& O'Sullivan  (2006) and the
mass-to-light ratio of a  12\,Gyr solar metallicity stellar population
from  the Bruzual  \&  Charlot (2003)  library  with an  exponentially
declining  SFR  history  with  $\tau=1$\,Gyr.   In  the  case  of  the
Hornschemeier et al.  (2005)  galaxies stellar masses are estimated by
applying the  method of section  \ref{sec_sed} to the  SDSS photometry
and using the $z$-band mass-to-light ratio.

\begin{figure*}
 \rotatebox{0}{\includegraphics[height=1\columnwidth]{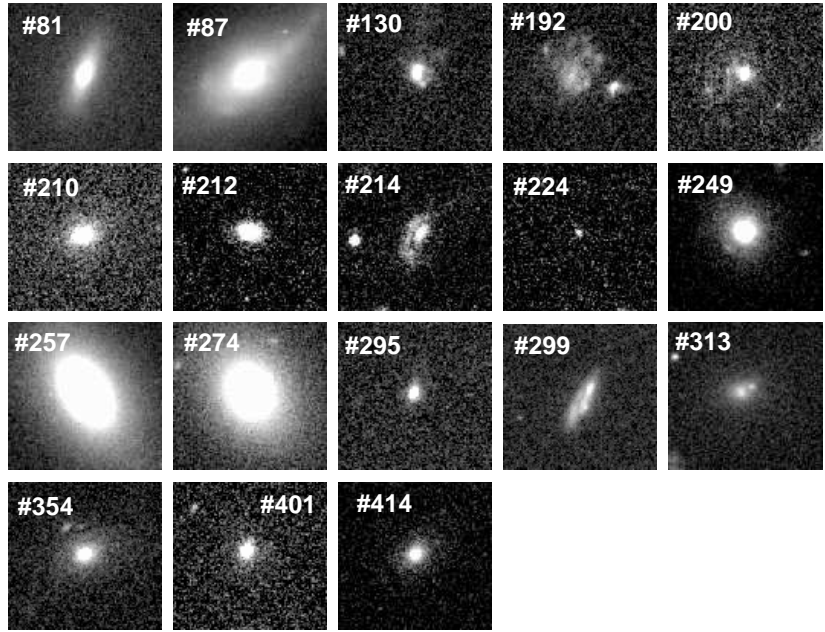}}
\caption{
  Hubble Space Telescope ACS V-band images of the infrared faint
  X-ray sources in the GOODS-North sample.
 }\label{fig_acs}
\end{figure*}

In Figure  \ref{fig_lxmass}, some of the CDF-N  sources are consistent
with  the  $M_{\star}-L_X$ relation  of  low-$z$ early-type  galaxies,
suggesting X-ray emission dominated by  hot gas and LMXRBs (i.e. \#81,
87, 210,  214, 257, 274,  313, 358, 401;  Table 1). For  those sources
that  optical spectra  are available  for visual  inspection,  we find
absorption  optical  lines only.   The  Hubble  Space Telescope  (HST)
Advanced   Camera  for   Surveys  (ACS)   images  also   suggest  E/S0
morphologies  for  most   of  these  systems  (Figure  \ref{fig_acs}).
Moreover, in Figure \ref{fig_lxmass} there are also many sources that
are  X-ray luminous  for their  stellar mass  and stand  out  from the
region  of the parameter  space occupied  by low-$z$  ellipticals (i.e
\#130,  192,  200,  212, 224,  249,  295,  299,  354, 414;  Table  1).
Low-luminosity  AGN activity  is an  obvious interpretation  for these
sources. However,  some of these  systems also show the  $\rm H\delta$
line  in absorption  with equivalent  width in  the  range 3--7\,\AA\,
(\#192,  200,  212 and  414),  superimposed  on  an absorption  and/or
emission  line  optical   spectrum.   Although  the  equivalent  width
uncertainties are  $\approx 30\%$, the features  above are reminiscent
of the K+A  or E+A class of galaxies (e.g. Dressler  \& Gunn 1992) and
suggest  post-starburst  systems. The  HST/ACS  optical morphology  of
some,  but not  all,  of  these sources  show  low surface  brightness
disturbances,  that  may  indicate past  interactions/mergers  (Figure
\ref{fig_acs}).   It is  therefore  possible that  these galaxies  are
observed just after a  star-formation episode, possibly triggered by a
merger  event. In  this scenario  the X-ray  emission is  dominated by
luminous  X-ray  binaries that  will  eventually  dim, bringing  these
systems      onto     the      elliptical     locus      of     Figure
\ref{fig_lxmass}. Post-starburst  galaxies could potentially represent
a large fraction of the  CDF-N sources which appear X-ray luminous for
their stellar  mass in Figure  \ref{fig_lxmass}.  This is  because for
many of  them we  cannot measure the  $\rm H\delta$ line.   Either the
optical spectra are not available in electronic form (\#224, 249, 354)
or  the $\rm  H\delta$  lies outside  the  observable spectral  window
(\#130).

\begin{figure}
 \rotatebox{0}{\includegraphics[height=1\columnwidth]{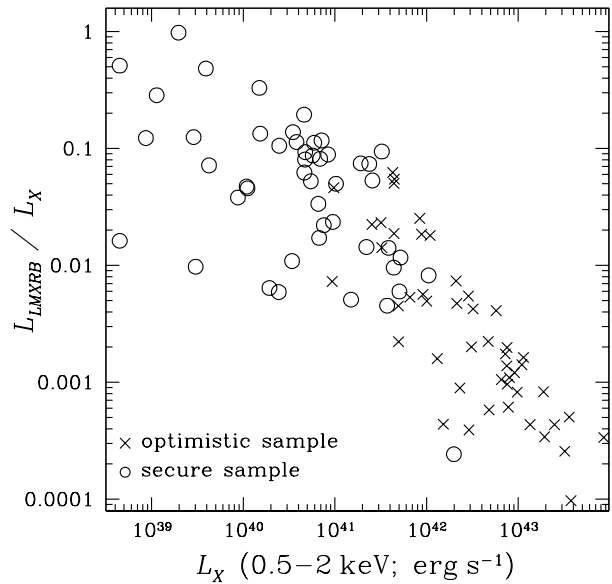}}
\caption{
 Fractional contribution of LMXRB to the total X-ray luminosity as a
 function of $L_X$. Only sources in the ``secure'' and ``optimistic''
 samples are plotted. The $L_{LMXRB}/L_X$ fraction is
 increasing at low X-ray luminosities, where quiescent galaxies with
 non-negligible LMXRB populations are found. 
 }\label{fig_frac}
\end{figure}

\begin{figure}
 \rotatebox{0}{\includegraphics[height=1\columnwidth]{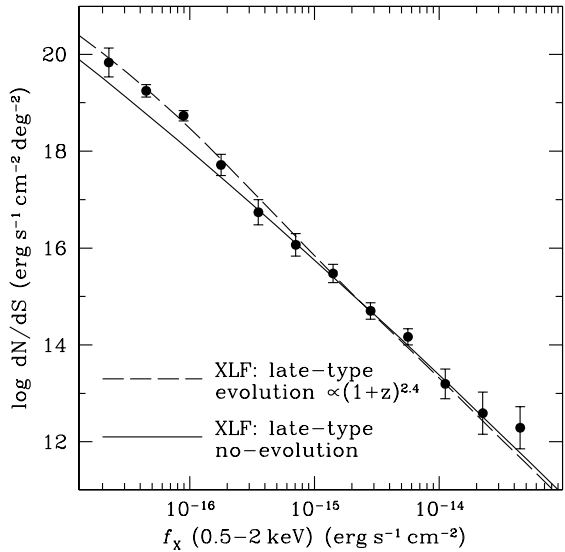}}
\caption{
 Differential ``normal galaxy'' counts in the 0.5-2\,keV spectral 
 band. The  continuous line is the prediction of the Georgakakis et
 al. (2006a) XLF for late type galaxies assuming no evolution. The
 dashed line is the prediction of the same XLF for the estimated maximum
 likelihood  evolution of the form $\propto (1+z)^{2.4}$. 
 }\label{fig_dnds}
\end{figure}

\section{Discussion}\label{sec_discussion}

\subsection{Sample selection}

Until  recently,   the  study   of  star-forming  galaxies   at  X-ray
wavelengths  was confined  to  the local  Universe  only via  targeted
observations of  few X-ray bright systems. This  was radically changed
by the {\it  Chandra} and the {\it XMM-Newton}  which have claimed the
identification,  for the  first time,  of X-ray  selected star-forming
galaxies at  cosmologically interesting redshifts opening  the way for
evolution studies (e.g. Hornschemeier et al. 2003; Norman et al. 2004;
Georgantopoulos et  al.  2005; Georgakakis  et al.  2006a).   The main
concern  however, is that  a fraction  of these  X-ray sources  may be
associated  with  weak  or  obscured  AGN and  therefore  their  X-ray
emission may be  due to accretion on a  supermassive black-hole rather
than stellar processes. The low X-ray--to--optical flux ratio does not
guarantee AGN free samples,  while optical spectroscopy often fails to
identify  AGN signatures  (e.g.  Moran  et al.  2002; Georgantopoulos,
Zezas \& Ward 2003; Peterson et al. 2006).

In  this paper  we address  the  AGN contamination  in X-ray  selected
galaxy  samples  by  exploiting  the  correlation  between  X-ray  and
infrared  luminosities  for star-forming  galaxies  (e.g.  Ranalli  et
al. 2003).  We use the  GOODS-North {\it Spitzer} data to estimate the
IR luminosity  of X-ray sources  in the CDF-North and  to discriminate
AGN from  star-formation dominated  systems. The effectiveness  of the
method is demonstrated in  Figure \ref{fig_lx_lir}.  Using this figure
we select a total of 45  sources that scatter $<2\sigma$ away from the
$L_X-L_{TOT}$  relation  for  local  star-forming  galaxies  (``normal
galaxy'' sample).   This limit  is chosen so  that the  final ``normal
galaxy''  sample  includes  sufficient  number of  sources  minimising
incompleteness and  AGN contamination. Changing  this selection cutoff
does not change our main conclusions.

This method however, although  efficient at high star-formation rates,
may be biased in the case  of quiescent galaxies.  This is because the
X-ray emission of starburst galaxies is dominated by HMXRBs which have
been  shown to  be good  tracers  of the  current star-formation  rate
(Grimm et  al.  2003; Persic et  al.  2004). In the  case of quiescent
systems however, the X-ray emission  is mainly from LMXRBs with a long
evolution  timescale. These  systems are  associated with  the stellar
mass  of  the host  galaxy  rather  than  the on-going  star-formation
activity. For example  Persic et al. (2004) argue  that correcting for
the LMXRB component significantly  reduces the scatter in the $L_X-\rm
SFR$ relation.   In Figure \ref{fig_lx_lir} we are  not accounting for
this effect and therefore we  might expect larger scatter for galaxies
in the low-luminosity end of  the plot, rendering our selection method
less reliable.  Nevertheless, the  tight linear relation between total
X-ray and infrared luminosities of  Ranalli et al.  (2003), which also
includes   quiescent  galaxies,  suggest   that  although   the  LMXRB
correction is important when calibrating  the $L_X$ as a SFR estimator
(e.g. Persic  et al. 2004), it  does not introduce biases  in the {\it
selection} of galaxies.

In   any  case,   we  quantify   the   effect  of   LMXRB  in   Figure
\ref{fig_lx_lir}  by estimating their  contribution to  the integrated
X-ray luminosity of  the galaxy. For this exercise  we use the results
from the study of Gilfanov (2004).  He found a linear relation between
the stellar  mass of nearby early-type galaxies,  $M_{\star}$, and the
total X-ray luminosity  of the X-ray point source  population of these
quiescent systems, expected to be dominated by LMXRB ($L_{LMXRB}$). We
adopt  the  mean  ratio  of  $L_{LMXRB}  /  M_{\star}  =  0.83  \times
10^{29}\rm \,  erg \, s^{-1}  M_{\odot}^{-1}$ found in this  study and
use the stellar masses estimated in section \ref{sec_sed} to determine
the  mean expected  LMXRB  X-ray  luminosity for  each  source in  our
sample.  The  fractional expected contribution  of LMXRB to  the total
X-ray luminosity is shown  in Figure \ref{fig_frac} for the ``secure''
and ``optimistic''  galaxy samples.  For  the majority of  the sources
this contribution is small ($<10\%$). This fraction however, increases
with decreasing  $L_X$, although there  are only a handful  of sources
for which $L_{LMXRB}$ is large enough, $>50\%$, to significantly alter
their position in  the $L_X-L_{TOT}$ diagram. Nevertheless, reapplying
the  selection  of ``normal  galaxies''  (i.e.   sources that  scatter
$<2\sigma$   away  from  the   $L_X-L_{TOT}$  relation)   using  X-ray
luminosities corrected for the  LMXRB contribution does not modify the
final ``normal  galaxy'' sample  used here. We  conclude that  for the
sample  studied here  the LMXRB  population has  little impact  on the
efficiency  of  Figure  \ref{fig_lx_lir}  for  selecting  star-forming
galaxies.

At the  bright-$L_X$ end of Figure \ref{fig_lx_lir},  we find evidence
for an upper limit in  the X-ray luminosity produced by star-formation
of $L_X\approx10^{42} \rm \, erg \, s^{-1}$. A similar result has been
reported  by Grimm  et al.   (2003) who  studied the  X-ray luminosity
function  of the  point-source population  (mostly X-ray  binaries) of
local late-type galaxies.  These authors  argue for a bright cutoff of
about few  times $10^{40}  \rm \, erg  \, s^{-1}$ in  the point-source
luminosity function of starbursts, with the highest luminosity systems
most likely  associated with HMXRBs.  Combining this  cutoff and their
finding that the normalisation  of the luminosity function scales with
SFR,  these authors  establish  a linear  relation between  integrated
X-ray luminosity  and SFR,  which implies that  extreme star-formation
rates are  required ($SFR \ga \rm  100 \, M_{\odot}$)  to produce $L_X
\ga 10^{42}  \rm \, erg \,  s^{-1}$. This is also  consistent with the
Ranalli  et al.   (2003)  relation in  Figure \ref{fig_lx_lir},  which
suggests that star-forming galaxies with  $L_X > 10^{42} \rm \, erg \,
s^{-1}$ should  have $L_{TOT}\sim  \rm 10^{46} \,  erg \,  s^{-1}$ and
therefore extreme SFRs.  Such objects however, may be too rare for the 
volume  probed  by the  CDF-North  to  be  detected. We  explore  this
possibility adopting the following  approach: we start from the 60$\mu
m$ LF  of cool infrared colour  galaxies presented by  Takeutchi et al.
(2003)  to  estimate the  expected  number  density  of sources  as  a
function of  redshift and  infrared luminosity.  Density  evolution of
the form $\propto (1+z)^{3.5}$,  proposed by these authors, is adopted
in this  calculation. We then use  the Ranalli et  al. (2003) relation
and its scatter to convert the infrared luminosity to $L_X$ in a given
redshift slice, under the assumption that the sources are dominated by
star-formation.   This  approach allows  us  to  estimate the  surface
number density of  starbursts with $L_X>10^{42} \rm \,  erg \, s^{-1}$
at different  redshift slices.   Convolving that with  the sensitivity
map of the  CDF-North we determine the expected  total number of these
systems within the surveyed area. We find that these sources are rare:
about  one  luminous  starburst  is expected  within  the  GOODS-North
area. We conclude that the size  of this sample is not large enough to
find this  class of  sources. Larger area  surveys at a  similar X-ray
depth are required to explore this issue. 

\subsection{X-ray evolution}

The identification of star-forming galaxies outside the local Universe
by the {\it Chandra} and  the {\it XMM-Newton} has prompted studies on
the  cosmological evolution  of this  population at  X-ray wavelengths
(e.g.   Norman et  al.  2004;  Ranalli  et al.   2005; Georgakakis  et
al. 2006b).   Although the star-formation history of  the Universe has
been mapped  out to $z\approx6$ by numerous  groups selecting galaxies
at  other wavebands  (e.g. Hopkins  2004), the  X-ray  regime provides
unique information on the physics of galaxies that is complementary to
any  other wavelength. Depending  on the  evolution timescales  of the
high and low  mass X-ray binary populations, which  dominate the X-ray
emission of star-forming  galaxies, models predict different evolution
rates  at  X-ray  wavelengths  (Ghosh  \&  White  2001).   Conversely,
constraining the  X-ray evolution of galaxies  can provide information
on physical properties, such  as the characteristic timescales of high
and low mass X-ray binaries.

We  quantify the X-ray  evolution of  star-forming galaxies  using the
``normal galaxy''  sample defined  in this paper  on the basis  of the
$L_X-L_{TOT}$ relation. This has  the advantage that is least affected
by AGN  contamination while  still including powerful  starbursts. The
median redshift of the sample is $z\approx0.5$. Although there are not
enough  sources (total  of 45)  to estimate  a luminosity  function at
different  redshift  bins  we  explore  the  X-ray  evolution  of  the
population by constructing the  differential counts $\rm dN/dS$. These
are  then  compared  with   the  predictions  of  the  observed  X-ray
luminosity   function   (XLF)   of   $z<0.1$   star-forming   galaxies
(Georgantopoulos  et  al.   2005;   Georgakakis  et  al  2006a)  under
different evolution  scenarios. At  the bright flux-end  we complement
the CDF-North  sources with  low-$z$ late-type galaxies  identified in
the  wide-area  shallow  surveys  summarised  by  Georgakakis  et  al.
(2006b).  In Figure \ref{fig_dnds}  we compare the differential counts
over 4\,dex in flux with  the $\rm dN/dS$ estimated by integrating the
late-type galaxy XLF of  Georgakakis et al. (2006a) parameterising the
evolution as  $\propto (1+z)^{p}$. In  this figure the $p=0$  curve is
inconsistent  with the  data.  A  maximum likelihood  fit yields  $p =
2.42^{+0.59}_{-0.66}$,  indicating  evolution  of this  population  to
$z\approx0.5$   similar   to  that   estimated   in  other   wavebands
(e.g. Hopkins 2004). This is  not surprising since a large fraction of
the X-ray selected ``normal galaxies'' used here (about 60\%) are LIRG
or ULIRGs  which are known to  evolve fast from the  local Universe to
higher-$z$ and are likely  dominated by HMXRBs. Finally, the evolution
rate  derived here,  $p  = 2.42^{+0.59}_{-0.66}$,  is consistent  with
those  models of Ghosh  \& White  (2001) that  adopt a  slow evolution
timescale  for the  low-mass  X-ray binary  population, $\ga  1$\,Gyr.
Shorter timescales predict higher exponents, almost independent of the
adopted functional form of the star-formation history of the Universe.

\subsection{Infrared-faint galaxy candidates}

In addition  to star-forming galaxies  our X-ray sample  also includes
infrared-faint sources. About 50\% of  them are early type systems with
X-ray  emission from  hot  gas and  LMXRBs.   The other  half of  this
population  deviates  from the  $M_{\star}-L_X$  relation for  low-$z$
early-type galaxies (see  Figure \ref{fig_lxmass}).  These sources are
likely   AGN   powered   but   optical  spectroscopy   also   suggests
post-starbursts. In this picture  the star-formation has stopped while
the X-ray emission  is dominated by luminous X-ray  binaries that will
eventually  dim, bringing  these systems  on the  elliptical  locus of
Figure \ref{fig_lxmass}.

Sipior  (2003) have developed  simulations of  the evolution  of X-ray
binaries   formed   in  a   burst   of   star-formation  lasting   for
20\,Myrs. Although their results  are for the $2-10$\,keV band instead
of the 0.5-2\,keV, their  main conclusions are nevertheless useful for
a  qualitative comparison  with the  observations presented  here.  In
this model  the 2-10\,keV luminosity  of the X-ray binaries  reaches a
maximum of $4\times10^{40} \rm \, erg \, s^{-1}$ shortly after the end
of  the star-formation burst.  At later  stages the  luminosity slowly
declines,  as new  populations of  lower-mass binaries  become active,
dropping  by about  1 and  2\,dex after  $\rm \approx0.1$  and 1\,Gyrs
respectively.  The normalisation  of the  light curve  depends  on the
stellar mass produced during the star-formation burst as well as model
dependent assumptions, such as  the evolution details of massive stars
($\rm M>20\,M_{\odot}$) during the  final stages of their lifetime and
the mass-loss through winds for main sequence stars. For example lower
mass-loss rates than those adopted  by Sipior (2003) will increase the
peak  luminosity by  a factor  of few  (e.g. Van  Bever  \& Vanbeveren
2000).  Assuming  that the  bulk  of  the  stellar population  in  the
infrared-faint sources in  our sample, about $\rm 10^{10}\,M_{\odot}$,
is produced during  a single burst of star-formation,  then scaling up
the  results  of  Sipior  (2003)  we estimate  an  instantaneous  peak
2-10\,keV  luminosity  of  $5\times10^{42}   \rm  \,  erg  \,  s^{-1}$
decreasing  to $5\times10^{41}  \rm  \,  erg \,  s^{-1}$  after a  few
hundred Myrs.

Read  \& Ponman  (1996) studied  the evolution  of the  X-ray emission
during merger-induced star-formation bursts.   They also find that the
$L_X/L_{B}$ ratio decreases from  interacting systems close to nuclear
coalescence (e.g.   ARP\,220) to merger  remnants, $\approx 1-2$\,Gyrs
after the burst (e.g.  NGC\,7252). At later stages ($\ga 4 \rm Gyrs$),
the X-ray luminosity  of elliptical galaxies is found  to increase, on
average, with  the age of their stellar  population presumably because
of the build-up of the  hot diffuse halo (O'Sullivan, Forbes \& Ponman
2001). The evidence above  suggests that if the infrared-faint sources
in the sample are post-starbursts  we are observing them shortly after
the star-formation has ended (about $0.1-1$\,Gyr).

At lower redshifts, $z \approx 0.1$, Hornschemeier et al. (2005) using
SDSS  data also  identified an  absorption-line X-ray  source  with no
optical  spectral evidence  for AGN  activity that  appears  too X-ray
luminous for  its stellar  mass (see Figure  \ref{fig_lxmass}).  These
authors also  discuss this  source in the  context of  sustained X-ray
emission following a burst  of star-formation. Such objects are likely
to be more  common at the high-$z$ Universe,  probed here, compared to
the present  day. We note however,  that the evidence in  favor of the
post-starburst scenario  is far from  conclusive.  For example  Yan et
al. (2006) argue that a large fraction of low-$z$ galaxies in the SDSS
with K+A  optical spectra  (i.e.  post-starbursts) show  weak emission
lines that are  consistent with Seyfert or LINER  type activity rather
than  residual  star-formation. Goto  (2006)  also  used  the SDSS  to
identify  systems with  deep  $\rm H\delta$ absorption  lines, typical  of
post-starbursts,  and optical  emission lines  typical of  AGN.  These
sources represent about 4\% of  the AGN population in a volume limited
subsample. Such  emission-line post-starburst galaxies  would not have
been selected  by traditional E+A/K+A criteria that  require little or
no  emission lines  to exclude  residual star-formation  activity. The
comparison between  the X-ray selected  sample presented here  and the
optically selected  SDSS post-starbursts, classified AGN  on the basis
of optical emission line  ratios, is not straightforward. For example,
the X-ray  properties of  the SDSS sources  remain unexplored.  At the
same  time the  optical emission  line properties  of the  CDF-N X-ray
sources (i.e. line  ratios) cannot be measured from  the existing data
because   of  the  lower   quality  spectra   compared  to   the  SDSS
(i.e. optically faint sources) and the different rest-frame wavelength
coverage because of the higher redshift.

\section{Conclusions}\label{sec_conclusions}

In this paper  we show that the $L_X  -L_{IR}$ relation established in
the local Universe is a powerful diagnostic for identifying starbursts
in  X-ray  surveys.   Using  this  approach we  compile  a  sample  of
moderate-$z$  ($z\approx0.5$) star-forming  galaxies in  the CDF-North
and conclude  that the observed  number density of this  population is
consistent  with X-ray  luminosity evolution  of the  form $(1+z)^{p}$
with  $p \approx  2.4$. This  rate of  evolution, is  similar  to that
inferred  from  other wavebands,  suggesting  that current  ultra-deep
X-ray samples are sensitive  enough to detect the starburst population
that drives the  global SFR evolution to $z\approx1$.   In addition to
star-forming  galaxies we  also  find a  population of  infrared-faint
passive galaxies that are however too X-ray luminous for their stellar
mass,  compared to  local ellipticals.   We  argue that  these may  be
galaxies  observed  $10^{8}-10^{9}$\,yr  after  the  end  of  a  major
star-formation  event  with  sustained  X-ray emission  due  to  X-ray
binaries produce during the burst. Over time, these sources are likely
to become fainter at X-ray wavelengths and will eventually evolve into
E/S0.   The present data  however, cannot  exclude the  possibility of
underlying  low-luminosity AGN  activity  in at  least  some of  these
sources.

\begin{table*} 
\caption{
The infrared-faint sample. 
}\label{tab1} 
\begin{center} 
\scriptsize
\begin{tabular}{lccc cccc cccc ccc}
\hline 
ID &
$\alpha_{X}$ & 
$\delta_{X}$ & 
$R$    &
$f_{3.6}$    & 
$f_{4.5}$    & 
$f_{5.8}$    & 
$f_{8.0}$    & 
$f_{24.0}$    & 
$\log L_X$ &
$\log L_{TOT}$ &
$\log M_{\star}$ &
$z$ & 
$\rm H\delta$ &
T\\
  &
(J2000)&
(J2000)&
(mag)&
($\mu$Jy) &
($\mu$Jy) &
($\mu$Jy) &
($\mu$Jy) &
($\mu$Jy) &
$(\rm erg \, s^{-1})$  &
($L_{\odot}$)  &
($M_{\odot}$)  &
  & 
 (\AA)& 
 \\
 (1) &
(2) &
(3)&
(4)&
(5) &
(6) &
(7) &
(8) &
(9) &
(10)  &
(11)  &
(12)  &
 (13) & 
 (14)& 
 (15)\\
\hline

81 & 12 36 08.06 & +62 08 52.4 & 20.28 & $81.91\pm0.06$ & $65.45\pm0.09$ & $47.34\pm0.49$ & $26.63\pm0.55$ & $<56.23$ & 40.79 &$-$ & 10.34 & $0.409^{}$ & $-$ & 3\\
87 & 12 36 09.75 & +62 11 45.9 & 18.52 & $94.45\pm0.06$ & $65.70\pm0.09$ & $43.11\pm0.48$ & $38.21\pm0.51$ & $<56.23$ & 39.81 &$-$ & 9.36 & $0.136^{A}$ & $-$ & 3\\
130 & 12 36 21.98 & +62 14 15.9 & 23.18 & $23.31\pm0.06$ & $23.33\pm0.09$ & $16.98\pm0.46$ & $17.88\pm0.49$ & $131.97\pm47.00$ & 41.44 &$-$ & 10.43 & $1.381^{EA}$ &$-$ &  3\\
192 & 12 36 36.25 & +62 13 19.2 & 22.64 & $6.40\pm0.06$ & $4.72\pm0.08$ & $3.68\pm0.46$ & $2.76\pm0.47$ & $<56.23$ & 40.82 &$-$ & 9.01 & $0.679^{EA}$ & $4$ & 3\\
200 & 12 36 37.35 & +62 08 31.1 & 23.06 & $28.82\pm0.08$ & $20.10\pm0.09$ & $16.64\pm0.48$ & $9.60\pm0.54$ & $<56.23$ & 41.31 &$-$ & 10.36 & $0.971^{EA}$ & $3$ & 3\\
210 & 12 36 39.72 & +62 15 47.3 & 22.53 & $82.92\pm0.06$ & $53.42\pm0.09$ & $41.89\pm0.48$ & $24.09\pm0.50$ & $<56.23$ & 41.51 &$-$ & 10.78 & $0.848^{}$ & $-$ & 3\\
212 & 12 36 40.12 & +62 16 55.6 & 23.36 & $35.79\pm0.06$ & $25.18\pm0.10$ & $18.93\pm0.47$ & $14.34\pm0.49$ & $<56.23$ & 41.77 &$-$ & 10.36 & $0.942^{A}$ & $7$& 1\\
214 & 12 36 40.54 & +62 18 33.2 & 23.55 & $61.20\pm0.06$ & $53.24\pm0.10$ & $38.36\pm0.50$ & $29.86\pm0.50$ & $<56.23$ & 41.37&$-$ & 11.22 & $1.144^{}$ & $-$ & 1\\
224 & 12 36 42.24 & +62 06 12.8 & 25.71 & $1.85\pm0.09$ & $1.52\pm0.15$ & $<2.82$ & $<3.55$ & $<56.23$ & 42.39 &$-$ & 9.05 & $1.070^{phot}$ & $-$ & 1\\
249 & 12 36 48.07 & +62 13 09.0 & 20.98 & $57.96\pm0.06$ & $43.37\pm0.08$ & $35.82\pm0.46$ & $21.59\pm0.48$ & $<56.23$ & 41.31 &$-$ & 10.17 & $0.475^{A}$ & $-$ & 3\\
257 & 12 36 49.45 & +62 13 47.1 & 18.47 & $172.86\pm0.06$ & $111.89\pm0.08$ & $74.03\pm0.46$ & $48.10\pm0.47$ & $<56.23$ & 38.98 &$-$ & 9.33 & $0.089^{A}$ & $-$ & 3\\
274 & 12 36 52.89 & +62 14 44.1 & 19.39 & $128.81\pm0.05$ & $104.68\pm0.09$ & $65.62\pm0.45$ & $44.37\pm0.47$ & $<56.23$ & 40.91 &$-$ & 10.62 & $0.320^{A}$ & $1.5$ & 3\\
295 & 12 36 57.36 & +62 10 25.5 & 23.83 & $16.23\pm0.06$ & $10.87\pm0.09$ & $8.34\pm0.47$ & $5.56\pm0.50$ & $<56.23$ & 41.32 &$-$ & 10.15 & $0.845^{EA}$ & $1.5$ & 1\\
299 & 12 36 57.91 & +62 21 28.6 & 22.43 & $21.51\pm0.10$ & $16.98\pm0.13$ & $<2.82$ & $<3.55$ & $<56.23$ & 42.34 &$-$ & 9.68 & $0.528^{A}$ & $<1$ & 1\\
313 & 12 37 01.10 & +62 21 08.1 & 21.76 & $121.74\pm0.07$ & $76.83\pm0.11$ & $64.09\pm0.57$ & $41.23\pm0.54$ & $<56.23$ & 41.31 &$-$ & 10.84 & $0.799^{A}$ & $<1$ &  3\\
354 & 12 37 08.68 & +62 15 01.5 & 22.17 & $21.40\pm0.05$ & $15.57\pm0.09$ & $12.34\pm0.45$ & $7.20\pm0.47$ & $<56.23$ & 41.07 &$-$ & 9.66 & $0.568^{A}$ & $-$ & 3\\
358 & 12 37 09.47 & +62 08 37.4 & 21.86 & $143.54\pm0.10$ & $100.55\pm0.16$ & $81.77\pm0.79$ & $56.54\pm0.90$ & $<56.23$ & 41.88 &$-$ & 10.97 & $0.907^{A}$ & $<1$ & 1\\
401 & 12 37 20.31 & +62 15 23.5 & 23.23 & $62.05\pm0.05$ & $48.83\pm0.09$ & $33.28\pm0.44$ & $21.46\pm0.45$ & $<56.23$ & 41.53 &$-$ & 10.95 & $0.936^{A}$ & $2$ & 1\\
414 & 12 37 24.43 & +62 12 41.8 & 22.18 & $37.59\pm0.06$ & $24.48\pm0.10$ & $19.63\pm0.47$ & $12.81\pm0.49$ & $<56.23$ & 41.46 &$-$ & 10.14 & $0.798^{EA}$ & $3$ & 3\\

\hline
\end{tabular} 

\begin{list}{}{}
\item The columns are: 1: identification number in the Alexander et al.
(2003) catalogue; 2: right ascension of the X-ray source; 3: declination 
of the X-ray source; 4: optical $R$-band magnitude; 5: $\rm 3.6\mu m$ 
flux in micro-Jy; 6: $\rm 4.5\mu m$ flux in micro-Jy; 7: $\rm 5.8\mu m$ 
flux in micro-Jy; 8: $\rm 8.0\mu m$ flux in micro-Jy; 9: $\rm 24\mu m$ 
flux in micro-Jy; 10: 0.5-2\,keV luminosity; 11: total IR luminosity in 
units of solar luminosity $\rm 3.83 \times 10^{33} \, erg \, s^{-1}$; 12: 
stellar mass in solar units; 13: redshift, the superscripts are 
``phot'': photometric redshift, ``A'': absorption line spectrum, ``EA'': 
emission+absorption line spectrum; 14: $\rm H\delta$ equivalent width 
in $\rm \AA$. Sources for which their spectra are not availabe for
visual inspection or the $\rm H\delta$ is outside the observable
spectal window have a dash;  15: X-ray source type  
from Bauer et al.  (2004), ``1'' is for ``optimistic''  sample and 
``3'' corresponds  to the ``secure'' sample.
\end{list}

\end{center}
\end{table*}

\begin{table*} 
\caption{
The star-forming candidate galaxy sample. 
}\label{tab2} 
\begin{center} 
\scriptsize
\begin{tabular}{lccc cccc cccc cc}
\hline 
ID &
$\alpha_{X}$ & 
$\delta_{X}$ & 
$R$    &
$f_{3.6}$    & 
$f_{4.5}$    & 
$f_{5.8}$    & 
$f_{8.0}$    & 
$f_{24.0}$    & 
$\log L_X$ &
$\log L_{IR}$ &
$\log M_{\star}$ &
$z$ & 
T\\
  &
(J2000)&
(J2000)&
(mag)&
($\mu$Jy) &
($\mu$Jy) &
($\mu$Jy) &
($\mu$Jy) &
($\mu$Jy) &
$(\rm erg \, s^{-1})$  &
($L_{\odot}$)  &
($M_{\odot}$)  &
 & 
 \\

 (1) &
(2) &
(3)&
(4)&
(5) &
(6) &
(7) &
(8) &
(9) &
(10)  &
(11)  &
(12)  &
 (13) &
(14)\\

\hline

46 & 12 35 55.43 & +62 15 05.0 & 18.99 & $70.72\pm0.10$ & $58.54\pm0.12$ & $43.01\pm0.76$ & $213.62\pm0.70$ & $371.39\pm8.52$ & 40.04 &9.52 & 9.47 & $0.132^{phot}$ & $3^{\star}$\\
56 & 12 35 58.94 & +62 15 42.4 & 20.26 & $26.83\pm0.12$ & $18.43\pm0.14$ & $15.75\pm0.87$ & $70.48\pm0.79$ & $202.24\pm7.13$ & 39.48 &8.63 & 8.24 & $0.076^{phot}$ & 3\\
57 & 12 35 59.72 & +62 15 50.0 & 19.10 & $152.87\pm0.12$ & $173.72\pm0.14$ & $129.30\pm0.91$ & $459.94\pm0.78$ & $1438.35\pm8.94$ & 40.77 &11.53 & 10.59 & $0.330^{phot}$ & $3^{\star}$\\
61 & 12 36 01.50 & +62 14 06.5 & 23.42 & $13.60\pm0.06$ & $14.54\pm0.09$ & $11.48\pm0.49$ & $11.19\pm0.52$ & $<56.23$ & 42.81 &12.60 & 10.45 & $2.033^{phot}$ & 1\\
67 & 12 36 03.31 & +62 11 11.0 & 20.72 & $133.54\pm0.06$ & $93.46\pm0.09$ & $112.38\pm0.48$ & $102.61\pm0.51$ & $1184.17\pm4.95$ & 41.40 &11.42 & 10.90 & $0.638^{}$ & $3^{\star}$\\
77 & 12 36 06.70 & +62 15 50.7 & 23.21 & $9.49\pm0.08$ & $13.07\pm0.10$ & $19.46\pm0.59$ & $25.42\pm0.57$ & $122.44\pm6.88$ & 43.55 &12.31 & 10.33 & $2.415^{}$ & 1\\
78 & 12 36 06.71 & +62 12 20.5 & 24.21 & $15.40\pm0.06$ & $9.76\pm0.09$ & $8.06\pm0.48$ & $6.47\pm0.51$ & $<56.23$ & 41.67 &9.36 & 10.31 & $0.747^{}$ & 1\\
80 & 12 36 07.68 & +62 15 03.3 & 24.34 & $19.05\pm0.06$ & $22.13\pm0.09$ & $18.42\pm0.48$ & $15.38\pm0.50$ & $36.15\pm5.77$ & 43.24 &10.06 & 10.68 & $1.443^{phot}$ & 1\\
83 & 12 36 08.22 & +62 15 53.1 & 21.51 & $18.67\pm0.08$ & $17.17\pm0.10$ & $15.17\pm0.57$ & $21.95\pm0.56$ & $128.16\pm6.56$ & 41.07 &10.13 & 9.45 & $0.458^{}$ & 3\\
94 & 12 36 12.03 & +62 09 00.4 & 25.08 & $17.10\pm0.06$ & $19.16\pm0.09$ & $19.87\pm0.47$ & $13.29\pm0.53$ & $<56.23$ & 42.59 &12.40 & 10.96 & $1.630^{phot}$ & 1\\
101 & 12 36 14.42 & +62 13 19.0 & 20.59 & $79.48\pm0.06$ & $62.93\pm0.09$ & $49.47\pm0.47$ & $55.18\pm0.50$ & $156.19\pm4.84$ & 40.83 &10.40 & 10.52 & $0.454^{}$ & $3^{\star}$\\
103 & 12 36 14.51 & +62 07 18.4 & 22.07 & $56.46\pm0.08$ & $43.72\pm0.10$ & $38.07\pm0.61$ & $33.42\pm0.62$ & $366.76\pm9.86$ & 41.96 &11.63 & 10.48 & $0.811^{phot}$ & $1^{\star}$\\
111 & 12 36 16.81 & +62 14 36.0 & 21.33 & $94.35\pm0.06$ & $64.22\pm0.08$ & $53.19\pm0.47$ & $37.36\pm0.48$ & $59.12\pm4.84$ & 40.65 &9.64 & 10.72 & $0.516^{}$ & 3\\
113 & 12 36 17.06 & +62 10 11.6 & 22.59 & $50.63\pm0.06$ & $36.27\pm0.08$ & $31.59\pm0.45$ & $25.13\pm0.51$ & $104.57\pm4.65$ & 42.75 &10.96 & 10.67 & $0.845^{}$ & 1\\
115 & 12 36 18.00 & +62 16 35.0 & 21.02 & $89.35\pm0.07$ & $83.77\pm0.10$ & $84.17\pm0.54$ & $94.93\pm0.56$ & $630.13\pm6.02$ & 42.99 &11.49 & 10.59 & $0.680^{}$ & 1\\
117 & 12 36 19.15 & +62 14 41.6 & 26.01 & $10.43\pm0.06$ & $12.56\pm0.08$ & $14.69\pm0.46$ & $16.17\pm0.48$ & $<56.23$ & 42.49 &10.05 & 10.21 & $0.770^{phot}$ & 1\\
119 & 12 36 19.45 & +62 12 52.4 & 20.58 & $99.90\pm0.06$ & $89.86\pm0.09$ & $87.73\pm0.46$ & $160.15\pm0.49$ & $935.80\pm4.84$ & 40.85 &11.20 & 10.69 & $0.473^{}$ & $3^{\star}$\\
124 & 12 36 20.99 & +62 14 12.2 & 24.51 & $12.06\pm0.06$ & $13.00\pm0.08$ & $12.66\pm0.46$ & $9.76\pm0.47$ & $<56.23$ & 41.95 &9.65 & 10.09 & $0.729^{phot}$ & 1\\
126 & 12 36 21.09 & +62 12 08.2 & 22.44 & $45.96\pm0.06$ & $31.32\pm0.08$ & $32.58\pm0.47$ & $25.66\pm0.49$ & $272.13\pm4.82$ & 41.01 &11.34 & 10.48 & $0.841^{}$ & $3^{\star}$\\
132 & 12 36 22.53 & +62 15 45.2 & 20.74 & $68.23\pm0.06$ & $48.33\pm0.09$ & $57.59\pm0.48$ & $67.12\pm0.50$ & $700.01\pm4.59$ & 41.08 &11.68 & 10.50 & $0.639^{}$ & $3^{\star}$\\
134 & 12 36 22.65 & +62 10 28.5 & 25.58 & $7.65\pm0.06$ & $9.89\pm0.09$ & $13.90\pm0.46$ & $26.22\pm0.52$ & $104.02\pm4.95$ & 42.23 &10.18 & 9.82 & $0.520^{phot}$ & 1\\
136 & 12 36 22.76 & +62 12 59.7 & 20.91 & $59.44\pm0.06$ & $45.48\pm0.08$ & $36.44\pm0.46$ & $28.86\pm0.48$ & $71.76\pm4.70$ & 40.67 &10.08 & 10.41 & $0.472^{}$ & 3\\
138 & 12 36 23.00 & +62 13 46.9 & 21.21 & $42.31\pm0.06$ & $39.50\pm0.08$ & $35.06\pm0.46$ & $73.23\pm0.47$ & $265.57\pm4.76$ & 40.82 &10.91 & 10.11 & $0.484^{}$ & $3^{\star}$\\
139 & 12 36 23.64 & +62 18 36.9 & 23.30 & $68.09\pm0.13$ & $51.52\pm0.19$ & $38.75\pm1.12$ & $33.54\pm1.05$ & $89.27\pm8.11$ & 42.61 &11.06 & 10.94 & $1.012^{}$ & 1\\
149 & 12 36 27.31 & +62 12 57.7 & 22.70 & $24.62\pm0.06$ & $22.69\pm0.08$ & $15.48\pm0.45$ & $17.75\pm0.47$ & $138.21\pm4.78$ & 41.63 &11.58 & 10.39 & $1.222^{}$ & $3^{\star}$\\
166 & 12 36 31.24 & +62 12 36.7 & 21.89 & $11.83\pm0.06$ & $11.29\pm0.08$ & $9.57\pm0.44$ & $17.76\pm0.46$ & $100.68\pm4.99$ & 40.53 &10.11 & 9.34 & $0.456^{}$ & $3^{\star}$\\
174 & 12 36 33.23 & +62 08 34.8 & 21.92 & $101.81\pm0.06$ & $93.58\pm0.09$ & $102.00\pm0.48$ & $118.68\pm0.54$ & $752.00\pm4.91$ & 43.03 &12.04 & 11.04 & $0.934^{}$ & 1\\
177 & 12 36 33.67 & +62 10 05.7 & 22.60 & $72.48\pm0.07$ & $56.94\pm0.09$ & $45.76\pm0.48$ & $48.24\pm0.53$ & $546.62\pm4.99$ & 42.14 &12.08 & 10.96 & $1.015^{}$ & 1\\
180 & 12 36 34.46 & +62 12 13.0 & 19.37 & $253.25\pm0.06$ & $208.47\pm0.08$ & $184.86\pm0.43$ & $319.99\pm0.46$ & $1265.15\pm4.93$ & 41.25 &11.29 & 10.92 & $0.457^{}$ & $3^{\star}$\\
182 & 12 36 34.50 & +62 12 41.2 & 23.32 & $62.42\pm0.06$ & $71.51\pm0.08$ & $58.82\pm0.44$ & $70.47\pm0.45$ & $458.54\pm4.94$ & 41.92 &12.25 & 11.09 & $1.219^{}$ & $1^{\star}$\\
194 & 12 36 36.75 & +62 11 56.0 & 22.32 & $21.89\pm0.06$ & $15.94\pm0.08$ & $13.97\pm0.43$ & $11.67\pm0.47$ & $<56.23$ & 42.31 &10.08 & 10.08 & $0.556^{}$ & 1\\
197 & 12 36 37.18 & +62 11 35.0 & 18.12 & $97.82\pm0.06$ & $69.95\pm0.08$ & $77.51\pm0.44$ & $374.34\pm0.47$ & $721.52\pm4.80$ & 38.90 &9.34 & 8.79 & $0.078^{}$ & $3^{\star}$\\
205 & 12 36 39.09 & +62 09 43.9 & 24.07 & $27.98\pm0.06$ & $27.78\pm0.09$ & $21.83\pm0.47$ & $20.22\pm0.52$ & $160.82\pm4.80$ & 42.96 &11.57 & 10.96 & $1.344^{phot}$ & 1\\
209 & 12 36 39.70 & +62 10 09.6 & 22.02 & $26.91\pm0.06$ & $22.06\pm0.08$ & $18.11\pm0.46$ & $27.41\pm0.50$ & $170.62\pm53.53$ & 41.38 &10.56 & 10.11 & $0.510^{}$ & 1\\
211 & 12 36 39.92 & +62 12 49.9 & 22.17 & $57.34\pm0.06$ & $40.69\pm0.08$ & $43.43\pm0.45$ & $32.93\pm0.47$ & $489.42\pm4.86$ & 41.36 &11.58 & 10.55 & $0.846^{}$ & $3^{\star}$\\
222 & 12 36 42.20 & +62 15 45.5 & 22.23 & $111.69\pm0.06$ & $95.62\pm0.09$ & $110.75\pm0.48$ & $127.69\pm0.49$ & $808.79\pm5.00$ & 42.33 &11.94 & 11.14 & $0.857^{}$ & 1\\
225 & 12 36 43.10 & +62 11 08.0 & 22.86 & $11.04\pm0.06$ & $13.35\pm0.08$ & $15.63\pm0.46$ & $17.18\pm0.48$ & $<56.23$ & 42.81 &12.51 & 10.61 & $3.234^{}$ & 1\\
227 & 12 36 44.00 & +62 12 50.1 & 21.36 & $50.44\pm0.06$ & $40.51\pm0.09$ & $49.12\pm0.45$ & $56.11\pm0.48$ & $417.88\pm4.97$ & 40.66 &11.01 & 10.23 & $0.556^{}$ & $3^{\star}$\\
230 & 12 36 44.40 & +62 11 33.3 & 22.51 & $151.06\pm0.06$ & $111.35\pm0.08$ & $77.98\pm0.46$ & $50.13\pm0.49$ & $<56.23$ & 41.63 &12.41 & 11.19 & $1.012^{}$ & $1^{\star}$\\
234 & 12 36 45.42 & +62 19 00.9 & 20.97 & $43.82\pm0.06$ & $41.34\pm0.10$ & $45.86\pm0.50$ & $88.27\pm0.50$ & $574.54\pm4.88$ & 40.87 &10.83 & 9.99 & $0.454^{}$ & $3^{\star}$\\
235 & 12 36 45.46 & +62 18 49.4 & 25.81 & $13.48\pm0.06$ & $19.02\pm0.10$ & $26.39\pm0.48$ & $40.99\pm0.49$ & $165.09\pm4.88$ & 43.45 &12.12 & 11.03 & $2.221^{phot}$ & 1\\
236 & 12 36 45.68 & +62 14 48.4 & 25.53 & $12.81\pm0.06$ & $18.42\pm0.09$ & $16.01\pm0.47$ & $24.52\pm0.50$ & $224.14\pm61.39$ & 41.50 &12.97 & 10.64 & $1.322^{phot}$ & 1\\
241 & 12 36 46.39 & +62 15 29.1 & 23.84 & $72.42\pm0.06$ & $52.30\pm0.10$ & $52.43\pm0.46$ & $40.63\pm0.49$ & $503.32\pm4.87$ & 41.83 &11.79 & 11.06 & $0.850^{}$ & $1^{\star}$\\
244 & 12 36 46.73 & +62 08 33.6 & 21.86 & $108.82\pm0.06$ & $82.13\pm0.09$ & $69.75\pm0.48$ & $64.18\pm0.54$ & $962.10\pm4.88$ & 41.37 &11.57 & 11.01 & $0.971^{}$ & $3^{\star}$\\
245 & 12 36 47.04 & +62 12 38.2 & 21.11 & $16.62\pm0.06$ & $14.03\pm0.08$ & $8.91\pm0.46$ & $16.02\pm0.49$ & $23.07\pm4.93$ & 39.92 &9.27 & 9.29 & $0.320^{}$ & 3\\
251 & 12 36 48.37 & +62 14 26.4 & 19.15 & $81.18\pm0.06$ & $59.03\pm0.08$ & $55.08\pm0.46$ & $350.69\pm0.47$ & $452.39\pm5.09$ & 39.62 &9.91 & 9.25 & $0.138^{}$ & $3^{\star}$\\
258 & 12 36 49.51 & +62 14 06.9 & 22.55 & $36.18\pm0.06$ & $23.89\pm0.08$ & $25.21\pm0.46$ & $18.66\pm0.47$ & $195.21\pm4.91$ & 40.97 &11.05 & 10.12 & $0.751^{}$ & $3^{\star}$\\

\hline
\end{tabular}
\end{center}
\end{table*}

\begin{table*}
\contcaption{}
\begin{center} 
\scriptsize
\begin{tabular}{lccc cccc cccc cc}
\hline

260 & 12 36 49.71 & +62 13 13.2 & 21.99 & $38.29\pm0.06$ & $39.51\pm0.08$ & $35.64\pm0.46$ & $94.18\pm0.48$ & $387.35\pm4.79$ & 40.73 &10.95 & 10.22 & $0.474^{}$ & $3^{\star}$\\
262 & 12 36 50.30 & +62 20 04.5 & 24.50 & $21.73\pm0.06$ & $17.28\pm0.10$ & $12.80\pm0.54$ & $10.94\pm0.52$ & $<56.23$ & 42.32 &9.77 & 10.56 & $0.811^{phot}$ & 1\\
264 & 12 36 51.03 & +62 17 31.8 & 22.33 & $7.05\pm0.06$ & $5.20\pm0.10$ & $3.01\pm0.46$ & $4.51\pm0.49$ & $<56.23$ & 40.28 &8.53 & 8.86 & $0.318^{}$ & 3\\
265 & 12 36 51.15 & +62 10 30.4 & 20.73 & $93.18\pm0.06$ & $98.17\pm0.09$ & $81.83\pm0.45$ & $273.38\pm0.49$ & $945.21\pm4.83$ & 40.76 &11.66 & 10.46 & $0.409^{}$ & $3^{\star}$\\
269 & 12 36 52.10 & +62 14 57.2 & 22.10 & $10.96\pm0.06$ & $11.86\pm0.09$ & $12.64\pm0.45$ & $13.50\pm0.47$ & $58.27\pm4.91$ & 40.38 &9.52 & 8.92 & $0.357^{}$ & 3\\
272 & 12 36 52.76 & +62 13 54.1 & 22.21 & $14.29\pm0.06$ & $14.69\pm0.08$ & $10.69\pm0.45$ & $11.54\pm0.48$ & $126.00\pm46.00$ & 41.57 &11.31 & 9.99 & $1.355^{}$ & $3^{\star}$\\
273 & 12 36 52.88 & +62 16 37.1 & 24.59 & $2.61\pm0.05$ & $3.81\pm0.10$ & $4.03\pm0.46$ & $5.27\pm0.49$ & $<56.23$ & 42.18 &11.83 & 9.59 & $1.937^{phot}$ & 1\\
277 & 12 36 53.37 & +62 11 39.6 & 22.93 & $33.50\pm0.06$ & $36.86\pm0.09$ & $30.47\pm0.48$ & $42.51\pm0.50$ & $319.76\pm4.90$ & 42.02 &12.17 & 10.70 & $1.275^{}$ &$3^{\star}$ \\
282 & 12 36 54.26 & +62 07 45.3 & 19.23 & $93.01\pm0.07$ & $81.84\pm0.12$ & $72.20\pm0.56$ & $445.25\pm0.71$ & $752.13\pm5.11$ & 39.05 &9.43 & 9.28 & $0.081^{phot}$ & $3^{\star}$\\
286 & 12 36 55.45 & +62 13 11.2 & 23.45 & $57.63\pm0.06$ & $41.43\pm0.09$ & $30.74\pm0.47$ & $21.28\pm0.50$ & $<56.23$ & 42.28 &11.30 & 10.90 & $0.955^{}$ & 1\\
287 & 12 36 55.79 & +62 12 00.9 & 24.71 & $15.60\pm0.06$ & $18.62\pm0.09$ & $22.34\pm0.48$ & $17.51\pm0.50$ & $202.26\pm4.83$ & 42.58 &12.36 & 10.87 & $2.061^{phot}$ & 1\\
288 & 12 36 55.89 & +62 08 07.6 & 22.54 & $114.45\pm0.07$ & $84.66\pm0.10$ & $80.56\pm0.53$ & $83.81\pm0.63$ & $820.86\pm4.88$ & 41.51 &11.74 & 11.25 & $0.792^{}$ &$3^{\star}$ \\
294 & 12 36 56.92 & +62 13 01.6 & 24.02 & $34.40\pm0.06$ & $29.05\pm0.09$ & $19.80\pm0.47$ & $16.07\pm0.50$ & $<56.23$ & 41.43 &9.46 & 10.43 & $0.474^{}$ & 1\\
300 & 12 36 58.33 & +62 09 58.5 & 17.92 & $215.85\pm0.06$ & $157.88\pm0.09$ & $122.58\pm0.46$ & $449.22\pm0.50$ & $523.22\pm4.84$ & 39.59 &9.90 & 10.04 & $0.137^{}$ & $3^{\star}$\\
305 & 12 36 58.85 & +62 16 37.9 & 20.00 & $66.23\pm0.06$ & $61.47\pm0.09$ & $42.97\pm0.46$ & $121.06\pm0.49$ & $263.72\pm4.82$ & 40.19 &10.27 & 10.08 & $0.298^{}$ & $3^{\star}$\\
309 & 12 36 59.80 & +62 19 33.9 & 24.02 & $31.40\pm0.05$ & $27.97\pm0.09$ & $19.62\pm0.45$ & $18.03\pm0.47$ & $<56.23$ & 42.69 &11.11 & 10.63 & $1.144^{}$ & 1\\
310 & 12 36 59.92 & +62 14 49.8 & 22.18 & $54.24\pm0.05$ & $36.96\pm0.09$ & $44.77\pm0.44$ & $35.33\pm0.46$ & $457.92\pm4.95$ & 40.92 &11.37 & 10.64 & $0.761^{}$ & $3^{\star}$\\
311 & 12 37 00.38 & +62 16 16.6 & 22.91 & $37.01\pm0.06$ & $27.92\pm0.09$ & $30.31\pm0.45$ & $27.55\pm0.48$ & $476.46\pm4.91$ & 41.34 &11.65 & 10.27 & $0.913^{}$ & $3^{\star}$\\
314 & 12 37 01.17 & +62 10 16.0 & 25.52 & $6.83\pm0.06$ & $8.61\pm0.09$ & $11.08\pm0.48$ & $9.64\pm0.50$ & $229.38\pm4.88$ & 42.00 &12.74 & 10.46 & $1.870^{phot}$ & $1^{\star}$\\
320 & 12 37 01.99 & +62 11 22.1 & 19.18 & $58.55\pm0.06$ & $42.76\pm0.09$ & $36.04\pm0.49$ & $164.37\pm0.50$ & $<56.23$ & 39.45 &9.64 & 9.32 & $0.136^{}$ & $3^{\star}$\\
322 & 12 37 02.62 & +62 12 43.9 & 25.42 & $10.29\pm0.06$ & $12.69\pm0.09$ & $15.06\pm0.48$ & $11.84\pm0.50$ & $68.00\pm4.94$ & 42.96 &9.39 & 10.53 & $2.341^{phot}$ & 1\\
327 & 12 37 03.65 & +62 11 22.8 & 23.46 & $10.00\pm0.06$ & $10.54\pm0.09$ & $13.10\pm0.49$ & $9.09\pm0.50$ & $158.16\pm4.91$ & 41.17 &11.27 & 9.65 & $0.913^{}$ & $3^{\star}$\\
333 & 12 37 04.64 & +62 16 51.9 & 20.60 & $39.36\pm0.05$ & $38.74\pm0.09$ & $30.24\pm0.45$ & $70.73\pm0.46$ & $187.69\pm4.98$ & 41.62 &10.37 & 9.81 & $0.376^{}$ & 1\\
339 & 12 37 06.12 & +62 17 11.9 & 18.73 & $139.63\pm0.05$ & $121.88\pm0.09$ & $81.29\pm0.44$ & $204.92\pm0.46$ & $636.25\pm5.03$ & 39.99 &10.27 & 10.35 & $0.253^{}$ & $3^{\star}$\\
345 & 12 37 06.96 & +62 08 30.1 & 25.05 & $19.39\pm0.09$ & $17.14\pm0.14$ & $10.77\pm0.70$ & $10.07\pm0.82$ & $<56.23$ & 42.33 &11.34 & 10.77 & $1.535^{phot}$ & 1\\
346 & 12 37 07.18 & +62 16 42.6 & 24.28 & $24.56\pm0.06$ & $18.71\pm0.09$ & $11.94\pm0.43$ & $9.08\pm0.46$ & $<56.23$ & 40.98 &9.82 & 10.42 & $1.016^{}$ & 1\\
349 & 12 37 07.56 & +62 19 56.1 & 25.94 & $19.62\pm0.06$ & $17.68\pm0.10$ & $12.47\pm0.46$ & $14.05\pm0.49$ & $67.04\pm4.93$ & 42.68 &11.17 & 10.71 & $1.249^{phot}$ & 1\\
351 & 12 37 07.70 & +62 16 16.0 & 22.79 & $77.88\pm0.05$ & $58.44\pm0.09$ & $44.99\pm0.44$ & $32.65\pm0.46$ & $215.01\pm4.88$ & 41.60 &11.37 & 10.98 & $0.940^{}$ & 1\\
353 & 12 37 08.33 & +62 10 55.9 & 20.39 & $60.90\pm0.06$ & $64.89\pm0.09$ & $52.89\pm0.49$ & $166.38\pm0.50$ & $700.07\pm4.86$ & 40.83 &11.46 & 9.83 & $0.422^{}$ & $3^{\star}$\\
359 & 12 37 09.68 & +62 08 41.1 & 22.74 & $114.14\pm0.10$ & $90.93\pm0.15$ & $90.14\pm0.77$ & $80.26\pm0.89$ & $1094.29\pm6.14$ & 41.65 &12.15 & 11.11 & $0.902^{}$ & $1^{\star}$\\
370 & 12 37 12.12 & +62 17 53.9 & 24.02 & $34.40\pm0.05$ & $34.15\pm0.09$ & $25.79\pm0.44$ & $29.68\pm0.46$ & $139.56\pm5.01$ & 42.77 &11.27 & 10.78 & $1.167^{phot}$ & 1\\
378 & 12 37 14.33 & +62 12 21.2 & 24.47 & $13.59\pm0.06$ & $11.65\pm0.10$ & $8.80\pm0.47$ & $7.33\pm0.49$ & $<56.23$ & 41.39 &10.77 & 10.52 & $1.084^{}$ & 1\\
381 & 12 37 14.77 & +62 16 17.0 & 24.02 & $17.52\pm0.05$ & $21.93\pm0.09$ & $19.48\pm0.42$ & $19.92\pm0.44$ & $124.58\pm4.86$ & 43.07 &13.00 & 10.80 & $1.522^{}$ & 1\\
383 & 12 37 15.94 & +62 11 58.3 & 18.36 & $254.71\pm0.06$ & $170.12\pm0.10$ & $119.67\pm0.47$ & $86.31\pm0.49$ & $213.05\pm4.86$ & 39.29 &9.05 & 10.05 & $0.112^{}$ & $3^{\star}$\\
387 & 12 37 16.19 & +62 11 30.5 & 24.80 & $26.49\pm0.06$ & $19.97\pm0.09$ & $12.52\pm0.47$ & $9.83\pm0.49$ & $<56.23$ & 41.64 &9.77 & 10.68 & $1.013^{phot}$ & 1\\
388 & 12 37 16.35 & +62 15 12.6 & 21.04 & $106.19\pm0.06$ & $76.06\pm0.09$ & $64.48\pm0.45$ & $56.03\pm0.47$ & $214.80\pm5.05$ & 40.19 &9.44 & 10.46 & $0.230^{}$ & 3\\
392 & 12 37 16.82 & +62 10 07.9 & 20.77 & $77.64\pm0.07$ & $70.25\pm0.10$ & $61.06\pm0.59$ & $140.57\pm0.68$ & $554.84\pm5.47$ & 40.54 &10.91 & 10.45 & $0.411^{}$ & $3^{\star}$\\
418 & 12 37 25.57 & +62 19 42.9 & 18.49 & $135.91\pm0.06$ & $133.29\pm0.10$ & $93.04\pm0.46$ & $358.98\pm0.49$ & $983.06\pm5.05$ & 40.39 &11.23 & 10.18 & $0.277^{}$ & $3^{\star}$\\
425 & 12 37 27.71 & +62 10 34.3 & 18.20 & $223.69\pm0.09$ & $185.50\pm0.14$ & $130.44\pm0.77$ & $507.10\pm0.95$ & $842.28\pm6.70$ & 38.65 &9.42 & 9.13 & $0.042^{phot}$ & $3^{\star}$\\
432 & 12 37 31.73 & +62 17 03.7 & 24.05 & $5.66\pm0.55$ & $7.77\pm0.09$ & $9.33\pm0.45$ & $17.32\pm0.47$ & $231.50\pm64.73$ & 42.70 &11.14 & 9.54 & $1.023^{phot}$ & 1\\
433 & 12 37 32.28 & +62 12 46.7 & 22.95 & $93.76\pm0.06$ & $72.34\pm0.10$ & $51.33\pm0.47$ & $38.69\pm0.50$ & $131.27\pm4.84$ & 41.65 &11.45 & 11.15 & $1.022^{}$ & $1^{\star}$\\
435 & 12 37 34.10 & +62 11 39.6 & 19.60 & $63.04\pm0.07$ & $51.51\pm0.19$ & $41.51\pm0.60$ & $208.75\pm0.72$ & $541.29\pm6.16$ & 40.03 &10.59 & 9.48 & $0.202^{}$ & $3^{\star}$\\
437 & 12 37 34.55 & +62 13 56.3 & 21.69 & $86.94\pm0.06$ & $61.17\pm0.09$ & $52.39\pm0.44$ & $38.00\pm0.46$ & $293.95\pm4.79$ & 42.46 &11.32 & 10.79 & $0.839^{}$ & 1\\
446 & 12 37 37.14 & +62 12 05.3 & 19.42 & $134.43\pm0.07$ & $124.27\pm0.13$ & $96.69\pm0.56$ & $189.78\pm0.67$ & $701.65\pm6.08$ & 40.59 &10.91 & 10.40 & $0.410^{}$ & $3^{\star}$\\
453 & 12 37 39.74 & +62 13 38.0 & 22.43 & $91.54\pm0.05$ & $64.34\pm0.09$ & $53.05\pm0.45$ & $38.19\pm0.47$ & $105.99\pm4.96$ & 42.11 &11.01 & 10.96 & $0.838^{}$ & 1\\
454 & 12 37 39.82 & +62 17 02.3 & 21.02 & $48.19\pm0.06$ & $36.15\pm0.10$ & $29.38\pm0.47$ & $22.43\pm0.49$ & $39.72\pm4.91$ & 41.49 &9.75 & 10.25 & $0.458^{}$ & 3\\
455 & 12 37 40.96 & +62 12 00.7 & 22.36 & $108.48\pm0.08$ & $126.47\pm0.17$ & $139.80\pm0.68$ & $179.12\pm0.89$ & $613.54\pm6.78$ & 43.72 &11.53 & 11.23 & $1.168^{}$ & 1\\
476 & 12 37 54.62 & +62 18 40.9 & 23.66 & $7.92\pm0.06$ & $5.51\pm0.11$ & $6.58\pm0.56$ & $5.39\pm0.56$ & $57.31\pm5.56$ & 40.97 &9.61 & 9.60 & $0.475^{}$ & 1\\
493 & 12 38 17.48 & +62 16 10.4 & 23.52 & $20.75\pm0.10$ & $18.96\pm0.23$ & $12.66\pm0.97$ & $14.03\pm1.08$ & $<56.23$ & 42.96 &12.52 & 10.81 & $1.454^{phot}$ & 1\\

\hline
\end{tabular} 
\begin{list}{}{}
\item The columns are: 1: identification number in the Alexander et al.
(2003) catalogue; 2: right ascension of the X-ray source; 3: declination 
of the X-ray source; 4: optical $R$-band magnitude; 5: $\rm 3.6\mu m$ 
flux in micro-Jy; 6: $\rm 4.5\mu m$ flux in micro-Jy; 7: $\rm 5.8\mu m$ 
flux in micro-Jy; 8: $\rm 8.0\mu m$ flux in micro-Jy; 9: $\rm 24\mu m$ 
flux in micro-Jy; 10: 0.5-2\,keV luminosity; 11: total IR luminosity in 
units of solar luminosity $\rm 3.83 \times 10^{33} \, erg \, s^{-1}$; 12: 
stellar mass in solar units; 13: redshift, the superscripts mark 
soources with photometric redshifts; 
14: X-ray source type from Bauer et al. (2004), ``1'' is for ``optimistic'' 
sample and ``3'' corresponds  to the ``secure'' sample. The $\star$
indicates sources in the ``normal galaxy'' samples, i.e. they scatter
$<2\sigma$  from  the X-ray/infrared  luminosity relation  of local
star-forming galaxies. dominated  by star-formation. 
\\
\end{list}
\end{center} 
\normalsize  
\end{table*}

\section{Acknowledgments}

We thank the anonymous referee for helpful comments that improved this
manuscript.  AG  acknowledges financial  support  from  PPARC and  the
Marie-Curie Fellowship grant  MEIF-CT-2005-025108.  This work is based
on  observations  made with  the  Spitzer  Space  Telescope, which  is
operated  by the  Jet Propulsion  Laboratory, California  Institute of
Technology under a contract with  NASA. We acknowledge use of the Team
Keck          Treasury         Redshift          Survey         (TKRS;
http://www2.keck.hawaii.edu/science/tksurvey/).


\begin{thebibliography}{}

\bibitem{1} Alexander D. M. et al.,  2003, AJ, 126, 539

\bibitem{2} Alexander D. M., Aussel H., Bauer F. E., Brandt W. N.,
Hornschemeier A. E., Vignali C., Garmire G. P., Schneider D. P., 2002,
ApJ, 568, 85L


\bibitem{3} Babbedge T. S. R., et al., 2006, MNRAS,  370, 1154


\bibitem{4} Babbedge, T. S. R, et al., 2004, MNRAS, 353, 654

\bibitem{5} Barger A. J., Cowie L. L., Wang W.-H., 2007, ApJ, 654, 764	

\bibitem{6} Barger, A. J. et al. 2003, AJ, 126, 632

\bibitem{7} Bauer, F.E., Alexander, D.M., Brandt, W.N., Schneider,
  D.P., Treister, E.,  Hornschemeier, A.E., Garmire, G.P., 2004, AJ,
  128, 2048  

\bibitem{8} Bauer F. E., Alexander D. M., Brandt W. N.,
  Hornschemeier A. E., Vignali C., Garmire G. P., Schneider D. P.,
  2002, AJ, 124, 2351

\bibitem{9} Belczynski K., Kalogera V., Rasio F. A., Taam R. E., Zezas A., Bulik
T., Maccarone T. J., Ivanova N., 2007, ApJS, submitted, astro-ph/0511811 

\bibitem{10} Belczynski K., Kalogera V., \& Bulik T., 2002, ApJ, 572, 407

\bibitem{11} Brandt W. N., Hornschemeier A. E., Schneider D. P., Alexander D. M.,
Bauer F. E., Garmire G. P., Vignali C., 2001, ApJ, 558L, 5	


\bibitem{12} Bruzual G., Charlot S., 2003, MNRAS, 344, 1000

\bibitem{13} Bundy K., Ellis R. S., Conselice C. J., 2005, ApJ, 625, 621

\bibitem{14} Capak P., et al., 2004, AJ, 127, 180

\bibitem{15} Chabrier G., 2003, ApJ, 586, 133L



\bibitem{17} Dressler A., \& Gunn J. E., 1992, ApJS, 78, 1

\bibitem{18} Dickinson, M., Giavalisco, M., and the GOODS Team, 2003,
  in "The Mass of Galaxies at Low and High Redshift," Proceedings of
  the ESO Workshop, Venice, Italy, 24-26 October 2001,
  eds. R. Bender \& A. Renzini, p. 324, astro-ph/0204213. 

\bibitem{19} Efstathiou A., Rowan-Robinson M., Siebenmorgen R., 2000,
  MNRAS, 313, 734  

\bibitem{20} Efstathiou A., Rowan-Robinson M., 1995, MNRAS, 273, 649

\bibitem{21} Ellis \& O'Sullivan, 2006, MNRAS, 367, 627


\bibitem{22} Fabbiano, G. 1995,  American Astronomical Society, 186th
  AAS Meeting, Bulletin of the American Astronomical Society, Vol. 27,
  p.877

\bibitem{23}  Georgakakis A. E., Chavushyan V., Plionis M.,
  Georgantopoulos I., Koulouridis E., Leonidaki I., Mercado A., 2006a,
  MNRAS, 367, 1017 

\bibitem{24} Georgakakis A., Georgantopoulos I., Akylas A., Zezas A.,
  Tzanavaris P., 2006b, ApJ, 641L, 101 

\bibitem{25} Georgakakis A., Georgantopoulos I., Stewart G. C., Shanks T.,
Boyle B. J., 2003, MNRAS, 344, 161

\bibitem{26} Georgantopoulos I., Zezas A., Ward M. J., 2003, ApJ, 584, 129 

\bibitem{27} Georgantopoulos I., Georgakakis A., Koulouridis E., 2005,
  ApJ, 624, 135

\bibitem{28} Ghosh P. \& White N. E., 2001, ApJ, 559, 97L

\bibitem{29} Gilfanov M., 2004, MNRAS, 349, 146

\bibitem{30} Grimm H.-J., Gilfanov M., Sunyaev R., 2003, MNRAS, 339, 793


\bibitem{31}  Gordon K. D., Clayton G. C., Misselt K. A., Landolt A. U.,
  Wolff M. J.,  2003, ApJ, 594, 279


\bibitem{32} Goto T., 2006, MNRAS, 369, 1765

\bibitem{33} Helou G., Khan I. R., Malek L., Boehmer L., 1988, ApJS, 68, 151

\bibitem{34} Hopkins, A., 2004, ApJ, 615, 209

\bibitem{35}  Hornschemeier A. E., Heckman T. M., Ptak A. F., Tremonti
C. A., Colbert E. J. M., 2005, AJ, 129, 86

\bibitem{36} Hornschemeier A. E. et al., 2003, AJ, 126, 575

\bibitem{37} Hornschemeier A. E., Brandt W. N., Alexander D. M., Bauer F. E.,
Garmire G. P., Schneider D. P., Bautz M. W., Chartas G., 2002, ApJ,
568, 82

\bibitem{38} Hornschemeier A. E., et al., 2000, ApJ, 541, 49


\bibitem{39} Kim D. W., et al., 2006, ApJ, 644, 829

\bibitem{39a} Laird E. S., Nandra K., Adelberger K. L., Steidel C. C.,
  Reddy N. A., MNRAS, 2005, 359, 47

\bibitem{39b} Laird E. S., Nandra K., Hobbs A., Steidel C. C., MNRAS,
  2006, 373, 217 

\bibitem{39c} Lilly S. J., Le Fevre O., Hammer F., Crampton D., 1996,
  ApJ, 460 ,1L

\bibitem{40} Moran E. C., Lehnert M. D., Helfand D. J. 1999, ApJ, 526, 649 

\bibitem{41} Moran E., Filippenko A. V., Chornock R., 2002, ApJ, 579L, 71

\bibitem{42} Nandra K., et al., 2002, ApJ, 576, 625

\bibitem{43} Norman C., et al., 2004, ApJ, 607, 721 

\bibitem{44} O'Sullivan E., Forbes Duncan A., Ponman Trevor J., 2001, MNRAS, 324,
420

\bibitem{45} Persic M., Rephaeli Y., Braito V., Cappi M., Della Ceca R.,
Franceschini A., Gruber D. E., 2004, A\&A, 419, 849


\bibitem{46} Peterson K. C., Gallagher S. C., Hornschemeier A. E., Muno M. P.,
Bullard, E. C., 2006, AJ, 131, 133


\bibitem{47} Ranalli P., Comastri A., Setti G., 2005, A\&A, 440, 23

\bibitem{48} Ranalli, P., Comastri, A. Setti, G., 2003, A\&A, 399, 39  

\bibitem{49} Read A. M., Ponman T. J., 2001, MNRAS, 328, 127

\bibitem{50} Rowan-Robinson M., et al., 2005, AJ, 129, 1183

\bibitem{51} Rowan-Robinson M., 2003, MNRAS, 345, 819

\bibitem{52} Rowan-Robinson M., 1995, MNRAS, 272, 737

\bibitem{53} Rowan-Robinson M., Benn C. R., Lawrence A., McMahon R. G.,
  Broadhurst T. J.,  1993, MNRAS, 263, 123

\bibitem{54} Sipior, M. S. 2003, PhD Thesis, The Pennsylvania State University

\bibitem{55} Takeuchi T. T., Buat V., Iglesias-P\'aramo J., Boselli A.,
  Burgarella D., 2005, A\&A, 432, 423 

\bibitem{56} Takeuchi T. T., Yoshikawa K., Ishii T. T., 2003, ApJ, 587L,
  89

\bibitem{57} Tzanavaris P., Georgantopoulos  I., Georgakakis A.,
  2006, A\&A, 454, 447 

\bibitem{58}  Zezas A. L., Georgantopoulos I., Ward M. J., 1998, MNRAS,
  301, 915 

\bibitem{59} Yan R., Newman J. A., Faber S. M., Konidaris N., Koo D., 
Davis M., 2006, ApJ, 648, 281

\bibitem{60}  Van Bever J., Vanbeveren D., 2000, A\&A, 358, 462

\bibitem{61} Verbunt F. \& Nelemans G. 2001, Class. Quantum Grav., 18, 4005 

\bibitem{61} Wilson G., Cowie L. L., Barger A. J., Burke D. J., 2002,
  AJ, 124, 1258

\end{thebibliography}
\end{document}